\documentclass{vldb}
\usepackage{balance}
\usepackage{times}
\usepackage{amsmath}
\usepackage{amssymb}
\usepackage{epsfig}
\usepackage[thmmarks]{ntheorem}
\usepackage{subfigure}
\usepackage{verbatim}
\usepackage{algorithmic}
\usepackage{multirow}
\usepackage{pbox}
\usepackage{url}
\PassOptionsToPackage{hyphens}{url}
\usepackage{hyperref}

\usepackage{microtype}
\makeatletter
\newif\if@restonecol
\makeatother

\usepackage[ruled, vlined]{algorithm2e}

\newcommand{\eat}[1]{}

\newcommand{\system}[1]{{\ensuremath {\mathsf{#1}}}\xspace}

\newcommand{\QED}{\mbox{\rule[0pt]{1.2ex}{1.2ex}}}
\qedsymbol{\QED}
\theoremsymbol{\QED}  

\theorembodyfont{\normalfont\upshape}
\theoremheaderfont{\normalfont\bfseries}

{
\theoremstyle{plain}
\newtheorem{definition}{Definition}

}

\newcommand{\entity}[1]{\textsf{\scriptsize #1}}

\newcommand{\etype}[1]{\textsc{\scriptsize #1}}
\newcommand{\edge}[1]{\textsf{\emph{\scriptsize #1}}}

\newcommand{\eg}{e.g.,\xspace}
\newcommand{\ie}{i.e.,\xspace}

\DeclareMathOperator*{\argmax}{arg\,max}

\title{Orion: Enabling Suggestions in a Visual Query Builder for Ultra-Heterogeneous Graphs}

\author{
{Nandish Jayaram\hspace*{0.9em}
Rohit Bhoopalam\hspace*{0.9em}
Chengkai Li\hspace*{0.9em}
Vassilis Athitsos
}\\
\fontsize{10}{10}\selectfont\itshape
University of Texas at Arlington
}

\begin{document}
\maketitle
\begin{abstract}
The database community has long recognized the importance of graphical query interface to the usability of data management systems.
Yet, relatively less has been done.
We present \system{Orion}, a visual interface for querying ultra-heterogeneous graphs.
It iteratively assists users in query graph construction by making suggestions via machine learning methods.
In its active mode, \system{Orion} automatically suggests top-$k$ edges to be added to a query graph.
In its passive mode, the user adds a new edge manually, and \system{Orion} suggests a ranked list of labels for the edge.
\system{Orion}'s edge ranking algorithm, Random Decision Paths (RDP), makes use of a query log to rank candidate edges by how likely they will match the user's query intent.
Extensive user studies using Freebase demonstrated that \system{Orion} users have a 70\% success rate in constructing complex query graphs, a significant improvement over the 58\% success rate by the users of a baseline system that resembles existing visual query builders.
Furthermore, using active mode only, the RDP algorithm was compared with several methods adapting other machine learning algorithms such as random forests and naive Bayes classifier, as well as class association rules and recommendation systems based on singular value decomposition.
On average, RDP required 40 suggestions to correctly reach a target query graph (using only its active mode of suggestion) while other methods required 1.5--4 times as many suggestions.
\end{abstract}

\vspace{-1mm}
\section{Introduction}\label{sec:intro}
The database community has long recognized the importance of graphical query interfaces to the usability of data management systems~\cite{lagunareport89}.
Yet, relatively less has been done and there remains a pressing need for investigation in this area~\cite{usability,beckmanreport14}.
Nevertheless, a few important ideas (e.g., Query-By-Example~\cite{qbe}) and systems (e.g., Microsoft SQL Query Builder) have been developed for querying relational databases~\cite{visual-tkde02}, web services~\cite{clide} and XML~\cite{xqbe,qursed}.

For querying graph data, existing systems~\cite{blau-etal-tr02-37, graphite, gblender, prague, vogue-cidr13, quble} allow users to build queries by visually drawing nodes and edges of query graphs, which can then be translated into underlying representations such as SPARQL and SQL queries.
While focusing on blending query processing with query formulation~\cite{graphite, gblender, prague, vogue-cidr13, quble}, existing visual query builders do not offer suggestions to users regarding what nodes/edges to include into query graphs.
At every step of visual query formulation, after adding a new node or a new edge into the query graph, a user would need to choose from a list of candidate \emph{labels}---names and types for a node or types for an edge.
The user, when knowing what label to use, can search the list of labels by keywords or sift through alphabetically sorted options using binary search.
But, oftentimes the user does not know the label due to lack of knowledge of the data and the schema.
In such a scenario, the user may need to sequentially comb the option list.
Furthermore, the user may not have a clear label in mind due to her vague query intent.

The lack of query suggestion presents a substantial usability challenge when the graph data require a long list of options, i.e., many different types and instances of nodes and edges.
The aforementioned systems~\cite{blau-etal-tr02-37, graphite, gblender, prague, vogue-cidr13, quble} were all deployed on relatively small graphs.
The crisis is exacerbated by the proliferation of \emph{ultra-heterogeneous graphs} which have thousands of node/edge types and millions of node/edge instances.
Widely-known ultra-heterogeneous graphs include Freebase~\cite{Bollacker+08freebase}, DBpedia~\cite{AuerBK+07}, YAGO~\cite{SuchanekKW07}, Probase~\cite{probase}, and the various RDF datasets in the ``linked open data''~\footnote{Linking open data. \url{http://www.w3.org/wiki/SweoIG/TaskForces/CommunityProjects/LinkingOpenData}.}.
Users would be better served, if graph query builders provided suggestions during query formulation.
In fact, query suggestion has been identified as an important feature-to-have among the desiderata of next-generation visual query interfaces~\cite{dbhci-dexa14}.

This paper presents \system{Orion}, a visual query builder that provides suggestions, iteratively, to assist users formulate queries on ultra-heterogeneous graphs.
\system{Orion}'s graphical user interface allows users to construct query graphs by drawing nodes and edges onto a canvas using simple mouse actions.
To allow schema-agnostic users to specify their exact query intent, \system{Orion} suggests candidate edge types by ranking them on how likely they will be of interest to the user, according to their relevance to the existing edges in the partially constructed query graph.
The relevance is based on the correlation of edge occurrences exhibited in a query log.
To the best of our knowledge, \system{Orion} is the first visual query formulation system that automatically makes ranked suggestions to help users construct query graphs.
The demonstration proposal for an early prototype of \system{Orion}~\cite{viiq} was based on a subset of the ideas in this paper.

\system{Orion} supports both an \emph{active} and a \emph{passive} operation mode.
(1) If the canvas contains a partially constructed query graph, \system{Orion} operates in the active
mode by default.
The system automatically recommends top-$k$ new edges that may be relevant to the user's query intent, without being triggered by any user actions.
Figure~\ref{fig:interface-viiq}(a) shows the snapshot of a partially constructed query graph, with nodes and edges suggested in the active mode.
The white nodes and the edges incident on them are newly suggested.
The user can select some of the suggested edges by clicking on them, and a mouse click on the canvas adds the selected edges to the partial query graph, and ignores the unselected edges.
(2) The passive mode is triggered when the user adds new nodes or edges to the partial query graph using simple mouse actions.
For a newly added edge, the suggested edge types are ranked based on their relevance to the user's query intent.
Figure~\ref{fig:interface-viiq}(c) shows the ranked suggestions for the newly added edge between the two nodes of types \etype{Person} and \etype{Film}, displayed in a pop-up box.
For a newly added node, labels are suggested for its type, the domain of its type, and its name if the node is to be matched with a specific entity.
The suggested labels are displayed in a pop-up box, as shown in Figure~\ref{fig:interface-viiq}(b), where type \etype{Person} is chosen as the label for the node.

The query construction process of a user can be summarized as a query session, consisting of positive and negative edges that correspond to edge suggestions accepted and ignored by the user, respectively.
At every step of the iterative process, based on the partially constructed query graph so far and the corresponding query session, \system{Orion}'s edge ranking algorithm---Random Decision Paths (RDP)---ranks candidate edges using a query log of past query sessions.
RDP ranks the candidate edges by how likely they will be of interest to the user, according to their correlation with the current query session's edges.
RDP constructs multiple decision paths using different random subsets of edges in the query session.
This idea is inspired by the ensemble learning method of random forests, which uses multiple decision trees.
Entries in the query log that subsume the edges of a decision path are used to find the ``support'' score of each candidate edge.
For each candidate, its support scores over all random decision paths are aggregated into its final score.
Section~\ref{sec:rcp} describes this ranking method in detail.
We also implemented several other edge ranking methods by adapting machine learning algorithms such as random forests (RF) and na{\"i}ve Bayes classifier (NB), as well as class association rules (CAR) and recommendation systems based on singular value decomposition (SVD).
Section~\ref{sec:baselineMethods} describes these techniques in detail.

To the best of our knowledge, there exists no publicly available real-world graph query log in the aforementioned form.
Existing visual query builders, possibly due to lack of users, do not have publicly available logs from their usage either.
The DBpedia SPARQL query benchmark~\cite{dbpedia-sparql} records queries posed by real users through the SPARQL query interface on DBpedia.
This can represent the positive edges in query sessions.
However, this query log may offer little help to \system{Orion}, due to two limitations: 1) It is applicable to DBpedia only and no other data graph, and 2) Only a third of the edge types present in DBpedia are used in the query log.
Hence, in addition to experimenting with this query log, we also simulated query logs for both Freebase and DBpedia data graphs using Wikipedia.
The premise is that the various relationships between entities, implied in the sentences of Wikipedia articles, represent co-occurring properties that simulate the positive edges in a query session.
Section~\ref{sec:workload} describes various ways of finding such positive edges and injecting negative edges, in order to simulate query logs.
Once \system{Orion} is in use, query sessions collected by it would result in a real-world query log that might be useful to the community in this line of research.

We conducted extensive user studies over the Freebase data graph, using 30 graduate students from the authors' institution, to compare \system{Orion} with a baseline system resembling existing visual query builders.
15 students worked on \system{Orion}, and the other 15 on the baseline system.
A total of 105 query tasks were performed by users of each system.
It was observed that \system{Orion} users had a 70\% success rate in constructing complex query graphs, significantly better than the 58\% success rate of the baseline system's users.
We also conducted experiments on both Freebase and DBpedia data graphs to compare RDP with other edge ranking methods---RF, NB, CAR and SVD.
The experiments were executed on the computing resources of the Texas Advanced Computing Center (TACC),~\footnote{\url{http://www.tacc.utexas.edu}.} to accommodate memory-intensive methods such as RF, SVD and CAR, which required between 40 GB to 100 GB of memory.
On average, the other methods required 1.5-4 times more suggestions to complete a query graph, compared to RDP's 40 suggestions.
The wall-clock time required to complete query graphs by RDP was mostly comparable with that of RF and NB, and significantly less than that of SVD and CAR.
We also performed experiments to study the effectiveness of the various query logs simulated.
RDP attained higher efficiency with the Wikipedia based query log compared to the query logs simulated using other ways discussed in Section~\ref{sec:workload}.\vspace{1mm}

We summarize the contributions of this paper as follows:
\vspace{-2mm}
\begin{list}{$\bullet$}
{ \setlength{\leftmargin}{0.7em} }
\item We present \system{Orion}, a visual query builder that helps schema-agnostic users construct query graphs by making automatic edge suggestions.
To the best of our knowledge, none of the existing visual query builders for graphs offers suggestions.
\vspace{-2mm}
\item  To help users quickly construct query graphs, \system{Orion} uses a novel edge ranking algorithm, Random Decision Paths (RDP), which ranks candidate edges by how likely they are to be relevant to the user's query intent.
RDP is trained using a query log containing past query sessions.
\vspace{-2mm}
\item There exists no such real-world query logs publicly available.
We thus designed several ways of simulating query logs.
Once \system{Orion} is in use, the real-world query log collected by it will become a valuable resource to the community.
\vspace{-2mm}
\item We conducted user studies on the Freebase data graph to compare \system{Orion}
with a baseline system resembling existing visual query builders.
\system{Orion} had a 70\% success rate of constructing complex query graphs, significantly better than the baseline system's 58\%.
\vspace{-2mm}
\item  We also performed extensive experiments comparing RDP with several other machine learning based methods, on the Freebase and DBpedia data graphs. Other methods required 1.5--4 times more suggestions than RDP, in order to complete query graphs.
\end{list}

\section{Related Work} \label{sec-related}

The unprecedented proliferation of linked data and large, heterogeneous graphs has sparked extensive interest in building knowledge-intensive applications.
The usability challenges in building such applications are widely recognized---declarative query languages such as SPARQL present a steep learning curve, as forming queries requires expertise in these languages and knowledge of data schema.
To tackle the challenges, a number of alternate querying paradigms for graph data have been proposed recently, including keyword search~\cite{blinks, freeq}, query-by-example~\cite{gqbe-icde14demo, gqbe-tkde, lim_edbt13, exemplarQueries}, natural language query~\cite{naturallang}, and faceted browsing~\cite{facet, facet1, facet2}.

Visual query builders~\cite{graphite, VISAGEiui15, gblender, prague, vogue-cidr13, quble} provide an intuitive and
simple approach to query formulation. Most of these systems deal with querying a
graph database and not a single large graph, except~\cite{quble,graphite, VISAGEiui15}. Firstly, it is unclear how
to directly apply the techniques proposed by systems that deal with graph databases to a single
large graph. This is because, their solutions work best on a data model with many small graphs,
rather than a single large graph. Secondly, these systems do not assist the user in query
formulation by automatically suggesting the new top-$k$ relevant edges.

\system{QUBLE}~\cite{quble}, \system{GRAPHITE}~\cite{graphite} and~\cite{VISAGEiui15} provide visual query interfaces
for querying a single large graph. But, they focus on efficient query processing,
and only facilitate query graph formulation by giving options to quickly draw various components
of the query graph.
Instead of recommending query components that a user might be interested in, they alphabetically
list all possible options for node labels (which may be extended to edge labels similarly).
They also deal with smaller data graphs. For instance, the graph considered by \system{QUBLE}
contains only around 10 thousand nodes with 300 distinct node types,
and they do not consider edge types. \system{Orion}, on the other hand, considers large graphs
such as Freebase, which has over 30 million distinct node types and 5 thousand distinct edge types.
With such large graphs, it is impractical to expect users to browse through all
options alphabetically to select the most appropriate edge to add to a query graph.
Ranking these edges by their relevance to the user's query intent is a necessity,
for which \system{Orion} is designed.

\section{System Overview}\label{sec:solution}
\vspace{-1mm}
\subsection{Data Model and Query Model}\label{sec:prelim}
An ultra-heterogeneous graph $G_d$, also called the data graph, is a connected, directed multi-graph with node set $V(G_d)$ and edge set $E(G_d)$.
A node is an entity~\footnote{Atomic values such as integers are not supported in the current version of the system.} and an edge represents a relationship between two entities.
The nodes and edges belong to a set of \emph{node types} $T_V$ and a set of \emph{edge types} $T_E$, respectively.
Each node (edge) type has a number of node (edge) instances.
Each node $v \in V(G_d)$ has an unique identifier, a name,~\footnote{Without loss of generality, we use a node's name as its identifier in presenting examples, assuming the names are unique.} and one or more node types $\mathrm{vtype}(v) \subseteq T_V$.
Each edge $e = (v_i,v_j) \in E(G_d)$, denoting a relationship from node $v_i$ to node $v_j$, belongs to a single \emph{edge type} $\mathrm{etype}(e) \in T_E$.

For example, \entity{Will Smith} and \entity{Tom Cruise} are instances of node type \etype{Film Actor}.
They are also instances of node type \etype{Person}.
There exist an edge (\entity{Tom Cruise}, \entity{Top Gun}) and another edge (\entity{Will Smith}, \entity{Men in Black}) which are both edges of type \edge{starring}.

The type of an edge constraints the types of the edge's two end nodes.
For instance, given any edge $e=(v_i,v_j)$ of edge type \etype{starring}, it is implied that $v_i$ is an instance of node type \etype{Film Actor} and $v_j$ is an instance of node type \etype{Film}.
In other words, \etype{Film Actor}$ \in \mathrm{vtype}(v_i)$ and \etype{Film} $\in \mathrm{vtype}(v_j)$.

\begin{figure}[t]
\centering
\includegraphics[width = 1.0\linewidth, keepaspectratio = true]{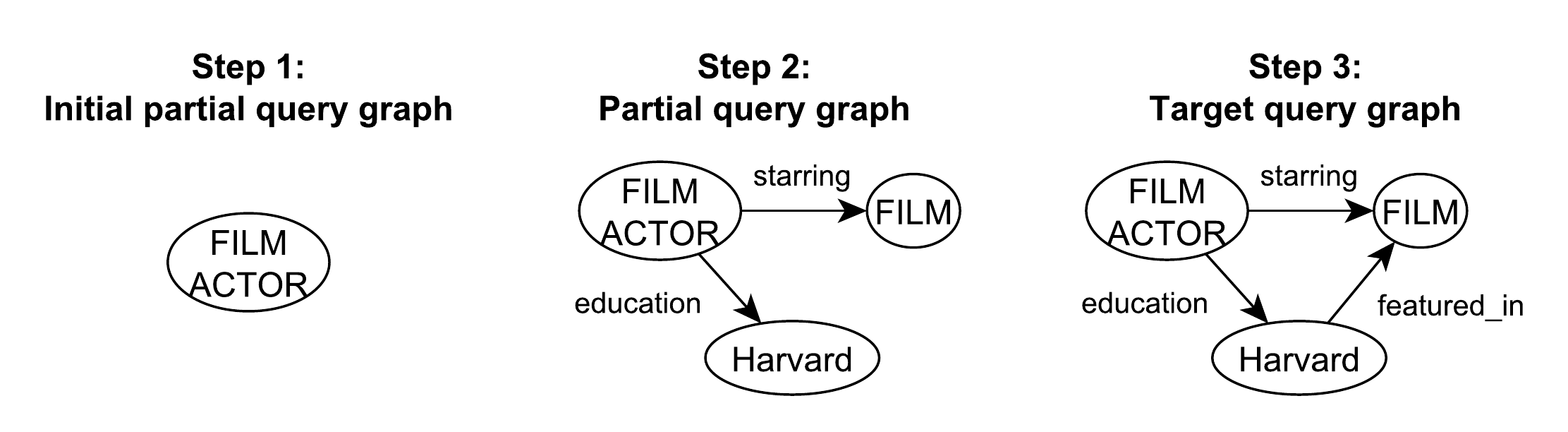}\vspace{-4mm}
\caption{Example Partial and Target Query Graphs}\vspace{1mm}
\label{fig:querygraphs}
\end{figure}

Given a data graph, users can specify their query intent through query graphs.
The concept of query graph is in Definition~\ref{def:qgraph}.
The nodes in a query graph are labeled by either names of specific nodes or node types.
Each answer graph to the query graph is a subgraph of the data graph and is edge-isomorphic to the
query graph.
In the answer graph, a node of the query graph is matched by a node of the specified name or any node of the specified type.
For instance, the query graph in Step 3 of Figure~\ref{fig:querygraphs} finds all \entity{Harvard}
educated film actors who starred in films featuring \entity{Harvard}.
In Figure~\ref{fig:querygraphs} and other query graphs in this paper, the all-capitalized node labels represent node types, while others represent node names.
\vspace{-1mm}

\begin{definition}[Query Graph]
\label{def:qgraph}
A query graph $G_q$ is a connected, directed multi-graph with node set $V(G_q)$ that may consist of both names
and types, and edge set $E(G_q)$, such that:\vspace{-2mm}
\begin{list}{$\bullet$}
{ \setlength{\leftmargin}{1em} \setlength{\itemsep}{-1pt} }
\item $V(G_q) \subseteq T_V \cup V(G_d)$.
\item $\forall e \in E(G_q), \mathrm{etype}(e) \in T_E$.
\end{list}
\end{definition}\vspace{-1mm}

\begin{figure*}[t]
\centering
\includegraphics[width = 0.83\linewidth, keepaspectratio = true]{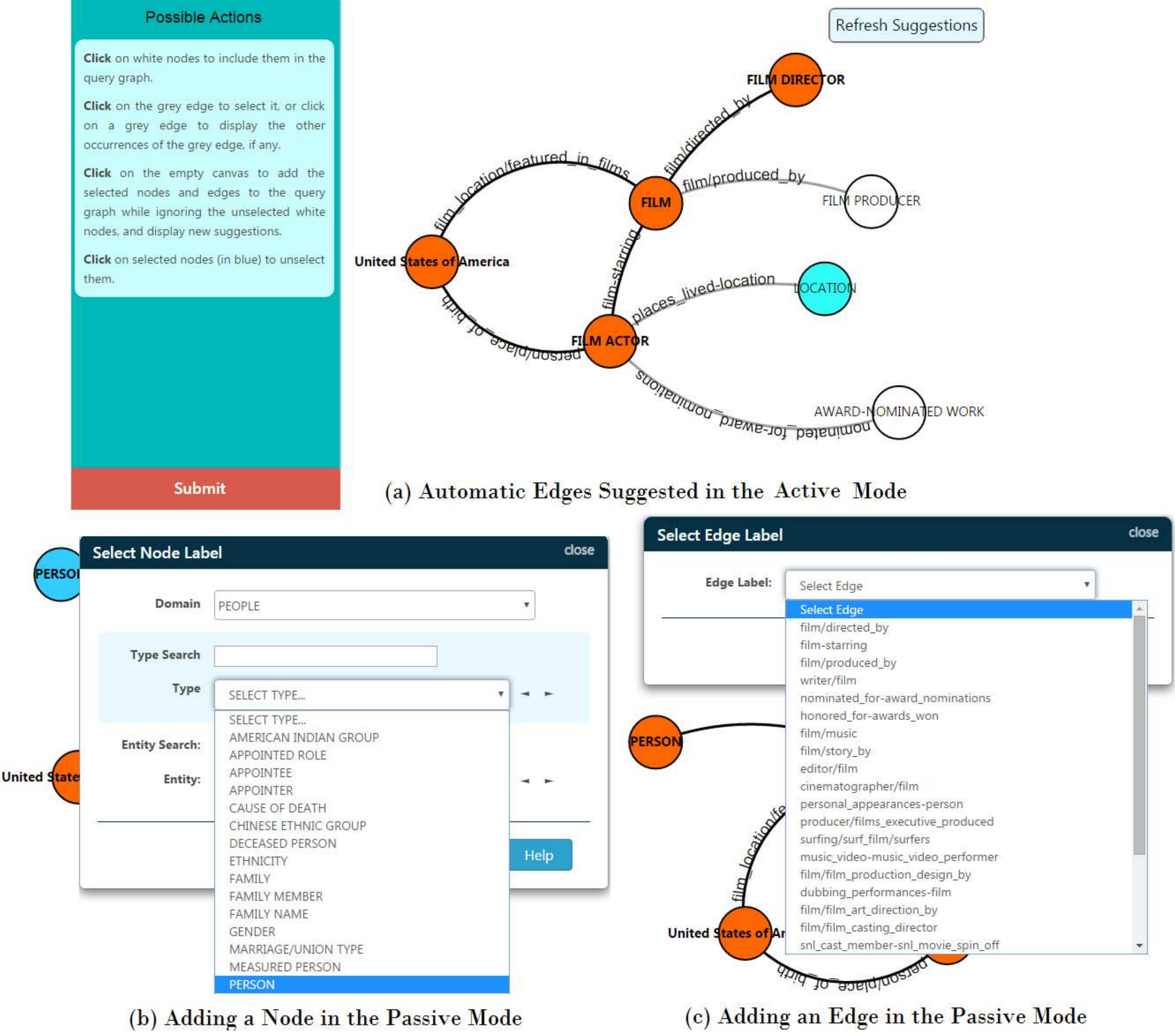}\vspace{-2mm}
\caption{User Interface of \system{Orion}}\vspace{-3mm}
\label{fig:interface-viiq}
\end{figure*}

\vspace{-1mm}
\subsection{User Interface for Providing Suggestions}\label{sec:ui-overview}

\system{Orion} helps users interactively and iteratively grow a partial query graph $G_p$ to a target query graph $G_t$.
It suggests edges to a user and solicit the user's response on the edges' relevance, in order to obtain a $G_t$ that satisfies the user's query intent.
The query session ends when either the user is satisfied by the constructed query graph or the user aborts the process.
The goal is to minimize the number of suggestions required to construct the target query graph.

Figure~\ref{fig:querygraphs} shows an example sequence of steps to construct a query graph.
The user starts by forming the initial partial query graph $G_p$ consisting of a single node.
Step 1 in Figure~\ref{fig:querygraphs} shows one such $G_p$ with a node of type \etype{Film Actor}.
New edges are then suggested to the user, who can choose to accept some of the suggestions.
For instance, step 2 in Figure~\ref{fig:querygraphs} shows the modified partial query graph obtained after adding two edges (together with two new nodes incident on the edges).
Without taking the suggested edges, the user can also directly add a new node or a new edge.
The system provides a ranked list of suggestions on the label of the new node/edge, for the user to choose from.
Step 3 in Figure~\ref{fig:querygraphs} shows the example target query graph obtained after adding the edge \edge{featured\_in} between \entity{Harvard} and \etype{Film}.
In general, to arrive at the target query graph $G_t$, the user continues the aforementioned process iteratively.
Figure~\ref{fig:interface-viiq}(a) shows the user interface of \system{Orion}.
It consists of a query canvas where the query graph is constructed.
In its active mode, \system{Orion} automatically suggests and displays top-$k$ new edges to add to the partial query graph.
In its passive mode, users use simple mouse actions on the query canvas to add new nodes and new edges.
\system{Orion} ranks candidate node and edge labels and displays them using drop-down lists in pop-up windows as shown in Figures~\ref{fig:interface-viiq}(b) and (c).
\system{Orion} also offers dynamic tips which list all allowable user actions at any given moment of the query construction process, as shown in Figure~\ref{fig:interface-viiq}(a).

\textbf{Active Mode:}\hspace{2mm}
An \system{Orion} user begins the query construction process by adding a single node into the empty canvas.
Once the canvas contains a partial query graph consisting of at least a node, \system{Orion} automatically operates in its active mode and suggests top-$k$ new edges.
Each suggested new edge is between two existing nodes or between an existing node and a new node.
Figure~\ref{fig:interface-viiq}(a) shows a partial query graph comprised of the four dark nodes and the edges between them.
The system suggests top-$3$ new edges, of which each is between an existing node (dark color) and a new node (white or light color).
The user can click on some white nodes (which then become light colored, e.g., \etype{Location} in Figure~\ref{fig:interface-viiq}(a)) to add them to the query graph, and ignore others.
The unselected white nodes are removed from display with a mouse click on the canvas, and the next set of new suggestions are automatically displayed.
If the user does not want to select any white nodes, a new set of suggestions can be manually triggered by clicking the ``Refresh Suggestions'' button on the query canvas.

\textbf{Passive Mode:}\hspace{2mm}
At any moment in the query construction process, a user can add a node or an edge using simple mouse actions, which triggers \system{Orion} to suggest labels for the newly added node/edge, i.e. it operates in the passive mode.
\textbf{1)} To add a new edge between two existing nodes in the partial query graph, the user clicks on one node and drags their mouse to the destination node.
The possible edge types for the newly added edge are displayed using a drop-down list in a pop-up suggestion panel, as shown in Figure~\ref{fig:interface-viiq}(c).
The edge types are ranked by their relevance to the query intent.
\textbf{2)} To add a new node, the user can click on any empty part of the canvas.
A suggestion panel pops up, as shown in Figure~\ref{fig:interface-viiq}(b).
It assists the user to select either a name or a type for the node.
The options in the two drop-down lists in Figure~\ref{fig:interface-viiq}(b), one for selecting names and the other for types, are sorted alphabetically.~\footnote{\system{Orion} currently ranks suggested edges by their relevance to users' query intent, in both active and passive modes.  How to rank node names/types based on query intent is an interesting future direction.}
To help the user find the desired node name or type, the suggestion panel is organized in a 3-level hierarchy.
Node types are grouped into domains.
The user can choose a domain first, followed by a node type in the domain and, if desired, the name of a specific node belonging to the chosen type.
The panel also allows the user to search for desired node name or type using keywords.
Right after the new node is added, it is not connected to the rest of the partial query graph.
\system{Orion} makes sure the partial query graph is connected all the time, except for such a moment.
Hence, no other operation is allowed, until the user adds an edge connecting the newly added node with some existing node, by using the aforementioned step 1).

\subsection{Candidate Edges}
\system{Orion} assists users in query construction by suggesting edge types to add to the partial query graph $G_p$, in both active and passive modes.
In its passive mode, a new edge is drawn between nodes $v$ and $v'$ by clicking the mouse on one node and dragging it to the other.
The set of candidate edges in the passive mode, $C_P$, consists of all possible edge types between $v$ and $v'$.
The set of candidate edges in the active mode, $C_A$, consists of any edge that can be incident on any node in $V(G_p)$, subject to the schema
of the underlying data graph.
A candidate edge can be either between two existing nodes in $G_p$, or between a node in $G_p$ and a new node automatically suggested along with the edge.

\vspace{-1mm}\begin{definition}[Incident Edges]\label{def:incidentEdges}
Given a data graph $G_d$, the incident edges $\mathrm{IE}(v)$ of a node $v \in V(G_d)$, is the set of types of the edges in $E(G_d)$ that are incident on node $v$. I.e., $\mathrm{IE}(v) = \{\mathrm{etype}(e) \lvert e=(v, v_i) \text{ or } e=(v_i, v), e \in E(G_d) \}$.
\end{definition}

\vspace{-1mm}\begin{definition}[Neighboring Candidate Edges]\label{def:incident-edges}
Given a partial query graph $G_p$, the neighboring candidate edges $\mathrm{NE}(v)$ of any node $v \in V(G_p)$, is the set of edge types defined as follows, depending on if $v$ is a specific node name or a node type (cf. Definition~\ref{def:qgraph}):\\
1) if $v \in V(G_d), \mathrm{NE}(v) = \mathrm{IE}(v)$;\\
2) if $v \in T_V, \mathrm{NE}(v) = \bigcup\{\mathrm{IE}(v') \lvert v' \in V(G_d) \text{, } v \in \mathrm{vtype}(v')\}$.
\end{definition}

When a new edge is added between two nodes $v$ and $v'$ in passive mode,
$C_P = \mathrm{NE}(v) \cap \mathrm{NE}(v')$, and the set of candidate edges in active mode is
$C_A = \bigcup_{v\in V(G_p)}\{e \lvert e\in  \mathrm{NE}(v) \}$.

\vspace{-1mm}\begin{definition}[Candidate Edges]\label{def:candidateEdges}
Candidate edges $C$ is the set of possible edges that can be added to the partial query graph
$G_p$ at any given moment in the query construction process.
\begin{multline}
\label{eq:candidateEdges}
	C\textsf{=}
	\begin{cases}
	C_P & \textsf{ in passive mode} \\
	C_A & \textsf{ in active mode}
	\end{cases}
\end{multline}
\end{definition}

In Section~\ref{sec:ranking} we discuss how to rank candidate edges and thus make suggestions to users in the query construction process.

\section{Ranking Candidate Edges}\label{sec:ranking}

A simple method to rank candidate edges is to order them alphabetically.
A more sophisticated method is to rank them by using statistics such as frequency in the data graph.
Such a method ignores information regarding users' intent.
A query log naturally captures different users' query intent.
It contains past query sessions which indicate what edges have been used together by users.
Such co-occurrence information gives evidence useful to rank candidate edges by their relevance to the user's query intent.

In a user's query session, edges found relevant, accepted and added to the query graph by the user
are called \emph{positive} edges.
In \system{Orion}'s active mode, suggested edges that are not accepted by the user are called \emph{negative} edges.
Both positive and negative edges play an important role in gauging the user's query intent, as evidenced by our experiments.
At any given moment in the query formulation process, the set of all positive and
negative edges hitherto forms a query session.

\vspace{-1mm}\begin{definition}[Query Log and Query Session]\label{def:querysession}
A query log $W$ is a set of query sessions.
A query session $Q$ is defined as a set of positive and negative edges.
$T_E$ (cf. Section~\ref{sec:prelim}) is the set of all possible positive edges for a data graph $G_d$.
The set of all possible negative edges, denoted $\overline{T_E}$, is defined as $\overline{T_E} = \cup_{e \in T_E} \{\overline{e}\}$.
If an edge $e \in T_E$ appears as a negative edge in a query session, it is represented as $\overline{e}$.
Let $T = T_E \cup \overline{T_E}$.
A query session $Q \in \mathcal{P}(T)$, where $\mathcal{P}(T)$ is the power set of $T$.
\end{definition}\vspace{-1mm}

Table~\ref{tab:querylog} shows an example query log containing 8 query sessions, one per line.
For instance, $w_4$ is a query session where the suggested edges $\overline{\edge{artist}}$ and $\overline{\edge{title}}$ were not accepted by the user, while edges \edge{writer} and \edge{director} were accepted.\vspace{-1mm}

{\flushleft \textbf{Problem Statement:}}\hspace{2mm}
Given a query log $W$, an ongoing query session $Q$ and a set of candidate edges $C$ (cf. Equation~\ref{eq:candidateEdges}), the problem is to rank the edges in $C$ by a scoring function that captures the likelihood that the user would find them relevant.

In Section~\ref{sec:baselineMethods}, we describe several baseline methods to rank candidate edges using query logs.
In Section~\ref{sec:rdp} we propose a novel method inspired by random forests.
Section~\ref{sec:workload} discusses several ways of obtaining a query log.

\subsection{Baseline Methods}\label{sec:baselineMethods}
Several machine learning algorithms can be adapted to rank candidate edges.
For instance, it can be seen as a recommendation problem.
One can also use a na{\"i}ve Bayes classifier or a random forest based classifier to find the probability that an edge $e$ is the \emph{class} associated with the ongoing query session $Q$, given by $P(e\lvert Q)$.
The query log $W$ can be used to learn such models off-line.
We implemented several baseline methods by adapting random forests (RF) and na¨ıve Bayes classifier (NB), as well
as class association rules (CAR)~\cite{car} and recommendation systems based on singular value decomposition
(SVD)~\cite{svd-reco}. Below we provide a brief sketch of these methods.

For RF and NB, we used a modified version of the query log $W$ as the training data.
A query session with $t$ positive edges and $t'$ negative edges was converted to $t$ training instances, with a different positive edge as the class of each training instance containing $t-1+t'$ attributes.
For instance, $w_1$ in Table~\ref{tab:querylog} was converted to $\langle (\edge{education},  \overline{\edge{nationality}}), (\edge{founder})\rangle$ and $\langle (\edge{founder},  \overline{\edge{nationality}}), (\edge{education})\rangle$, where \edge{founder} is the class of the first instance and \edge{education} the class for the second instance.
Multi-class classification models were learnt for RF and NB, wherein the number of classes equals the number of distinct positive edge types found in $W$.

For CAR, $W$ was modified to generate multiple rules.
The query sessions in $W$ are itemsets.
For a query session with $t$ positive edges and $t'$ negative edges, we generated $t$ association rules.
The antecedent (left hand side) of each rule contains $t-1+t'$ attributes, while the consequent (right hand side) contains exactly one positive edge.
For instance, $w_1$ in Table~\ref{tab:querylog} was converted to rules $\langle \edge{education},  \overline{\edge{nationality}} \rightarrow \edge{founder}\rangle$ and $\langle  \edge{founder},  \overline{\edge{nationality}} \rightarrow \edge{education}\rangle$.
If the antecedent of a rule and the ongoing session $Q$ overlap, the rule's consequent can be suggested to the user, weighted by the degree of overlap together with the commonly used measures of support and confidence in association rule mining.

For SVD, $W$ was converted to a $\lvert W \lvert$ rows $\times$ $\lvert T \lvert$ columns matrix.
Each element in the matrix was assigned a value of 0 or 1, based on their occurrence in the corresponding query session.
For example, for query log $W$ in Table~\ref{tab:querylog}, in the first row of the matrix, the columns corresponding to $\edge{education}$, $\edge{founder}$ and $\overline{\edge{nationality}}$ were set to 1, while the rest were set to 0.

\begin{table} [bt]
\centering
\scriptsize
\begin{tabular}{|p{6mm}|l|}
  \hline
  {\bf Id}  &   {\bf Query Session}\\  \hline\hline
  $w_1$ & \edge{education}, \edge{founder}, $\overline{\edge{nationality}}$ \\
  $w_2$ & \edge{starring}, $\overline{\edge{music}}$, \edge{director} \\
  $w_3$ & \edge{nationality}, $\overline{\edge{education}}$, \edge{music}, $\overline{\edge{starring}}$ \\
  $w_4$ & $\overline{\edge{artist}}$, $\overline{\edge{title}}$, \edge{writer}, \edge{director} \\
  $w_5$ & $\overline{\edge{director}}$, \edge{founder}, \edge{producer} \\
  $w_6$ & \edge{writer}, $\overline{\edge{editor}}$, \edge{genre} \\
  $w_7$ & $\overline{\edge{award}}$, \edge{movie}, \edge{director}, $\overline{\edge{genre}}$ \\
  $w_8$ & \edge{education}, \edge{founder}, $\overline{\edge{nationality}}$ \\
  \hline
  \end{tabular}
\caption{Example Query Log $W$}
\label{tab:querylog}
\end{table}

\subsection{Random Decision Paths (RDP)}
\label{sec:rdp}

Here we describe random decision paths (RDP), a novel method for measuring the relevance of a candidate edge. The RDP formulation is motivated by random forests~\cite{breiman_ml2001}. However, RDP has important differences from the standard definition and application of random forests, and significantly outperforms standard random forests in our experiments.

\subsubsection{Motivation: from Random Forests to Random Decision Paths}\label{sec:rdp_motivation}

To better understand the similarities and differences between RDP and random forests, it is useful to briefly review decision trees and random forests. In a general classification setting, a decision tree $D$ defines a probability function $P_D(y | x)$, where $x$ is a pattern, and $y$ is the class of that pattern. The decision tree $D$ can also be seen as a classifier that maps patterns to classes: $D(x) = \argmax_{y}P(y | x)$. The output of tree $D$ on a pattern $x$ is computed by applying to $x$ a test defined at the root of $D$, and using the result of the test to direct $x$ to one of the children of the root. Each child of the root is a decision tree in itself, and thus $x$ moves recursively along a path from the root to a leaf, based on results of tests applied at each node. A leaf node L stores precomputed probabilities $P_L(y)$ for each class y. If pattern $x$ ends up on a leaf $L$ of $D$, then the tree outputs $P_D(y | x) = P_L(y)$.

A random forest $F$ is a set of decision trees. A forest $F$ defines a probability $P_F(y | x)$, as the average $P_D(y | x)$ over all trees $D \in F$. To construct a random forest, each tree is built by choosing a random feature to test at each node, until reaching a predetermined number of trees. The probability values stored at the leaves of each tree are computed using a set of training patterns, for each of which the true class is known.

Random forests can be applied to our problem, but have certain undesirable properties. Each pattern is a query session, consisting typically of a few (or a few tens of) positive and negative edges. The total number of edge types can reach thousands (it equals 5253 in one of our experimental datasets). The test applied at each node of a decision tree simply checks if a certain edge (positive or negative) is present in the query session. Since query sessions contain relatively few edges compared to the number of edge types, for most tests the vast majority of results is a ``no'', meaning that the query session does not contain the edge specified in the test. This leads to highly unbalanced trees, where the path corresponding to all ``no'' results gets the majority of training examples, and paths corresponding to more than 1-2 ``yes'' results frequently receive no training examples. At classification time, the input pattern $x$ ends up at the all-no path most of the times, and thus the class probabilities $P_D(y | x)$ do not vary much from the priors $P(y)$ averaged over all training examples.

Our solution to this problem is mathematically equivalent to constructing a random forest on the fly, given a query session $Q$ to classify. This random forest is explicitly constructed to classify $Q$, and is discarded afterwards; a new forest is built for every $Q$. The tests that we use for tree nodes in that forest consider exclusively edges that appear in $Q$. This way, the probabilities stored at leaf nodes are computed from training examples that are similar to $Q$ in a sense, as they share at least some edges with $Q$. This is why we expect these probabilities to be more accurate compared to the probabilities obtained from a random forest constructed offline, without knowledge of $Q$. This expectation is validated in the experimental results.

At the same time, since we know $Q$, constructing full random forests is not necessary, and we can save significant computational time by exploiting that fact. The key idea is that, for any decision tree $D$ that we may build, since we know $Q$, we know the path that $Q$ is going to take within that tree. Computing the output for any other paths of $D$ is useless, since $D$ is constructed for the sole purpose of being applied to $Q$. Therefore, out of every tree in the random forest, we only need to compute and store a single path. Consequently, our random forest is reduced to a set of decision paths, and this set is what we call ``random decision paths'' (RDP).

\subsubsection{Formulation of Random Decision Paths}\label{sec:rcp}
We measure the relevance of a candidate edge $e$ to query session $Q$, by aggregating the
relevance of $e$ to several different subsets of edges in $Q$. We estimate the relevance of an
edge $e$ to each such subset of $Q$ using the query log $W$. We define a support function
$\mathrm{supp}(e, Q_i, W)$ to estimate the relevance of an edge $e$ to $Q_i \subseteq Q$:
\begin{align}
\label{eq:support}
\mathrm{supp}(e, Q_i, W) = \frac{\lvert \{w \lvert w\in W \textsf{, } Q_i \cup \{ e\} \subseteq w \}\lvert }
{\lvert \{w \lvert w\in W \textsf{, } Q_i \subseteq w \}\lvert }
\end{align}
The intuition behind using multiple subsets of $Q$ to measure the relevance of an edge $e$ to
the query session $Q$, instead of using the entire query session $Q$ alone is the following:
if $Q$ is long, \ie the query session contains a large number of positive and negative edges,
$\mathrm{supp}(e, Q_i, W)$ might be equal to 0 for every candidate edge $e$. This is because
it is unlikely to find any query session in the query log that is a super-set of $Q$.

If $\mathcal{P}(Q)$ is the power set of query session $Q$, we propose to build a set of random
decision paths $\Re$, that is: 1) a set of decision paths based only on the edges in
query session $Q$, and 2) a subset of $\mathcal{P}(Q)$ such that
$\lvert \Re \lvert \ll \lvert \mathcal{P}(Q) \lvert$.
We do not attempt to pre-learn a set of decision paths using query log $W$ that are used to
rank edges for any arbitrary query session (like learning a decision tree or rules for a classification model).
Instead, given a query session $Q$, we only build random decision paths specific to $Q$, that
measure the correlation of a candidate edge $e$ with different random subsets of edges in $Q$. In
other words, we assume the presence of a virtual space of all possible decision paths, but
only instantiate and use a few random paths specific to $Q$.

\begin{definition}[Decision Path] \label{def:correlationPath}
A decision path $\overrightarrow{O}$ is an ordered sequence of edges, for a set of edges $O$.
\end{definition}

The positive and negative edges in a query session $Q$ reflect the relevance and irrelevance
of the edges to the user's query intent. An example order for the decision path $\overrightarrow{Q}$
corresponding to query session $Q$ is the order of the edge suggestion sequence. There can be several
such ordered sequences for a query session. For any query session $O \in \mathcal{P}(T)'$,
the number of possible orders are equal to the total number of permutations of $O$, which is equal to
$\lvert O \lvert !$. Given the set of all query sessions $\mathcal{P}(T)'$, we define
$\overrightarrow{\mathcal{P}(T)'}$ as the set of all possible decision paths.
$\overrightarrow{\mathcal{P}(T)'} = \bigcup_{O \in \mathcal{P}(T)'}\{\overrightarrow{O_i} \lvert \forall i, 1 \leq i \leq \lvert O \lvert !\}$, and
$\lvert \overrightarrow{\mathcal{P}(T)'} \lvert$ is prohibitively large in practice.

A decision path $\overrightarrow{O}$ has a prefix path associated with it.
For instance, the prefix of a decision path $\overrightarrow{O}$, denoted by
$\mathrm{prefix}(\overrightarrow{O})$,
is the path before adding the last edge that formed $\overrightarrow{O}$. If
$\overrightarrow{O} = \{e_1, e_2, \ldots, e_{k-1}, e_k \}$, then
$\mathrm{prefix}(\overrightarrow{O}) = \{e_1, e_2, \ldots, e_{k-1}\}$.
The support for a decision path
$\overrightarrow{O}$ is given by $\mathrm{count}(\overrightarrow{O})$, defined as \vspace{-1mm}
\begin{align}
W_{\overrightarrow{O}} = \{w \lvert w \in W, O \subseteq w\},
\mathrm{count}(\overrightarrow{O}) = \lvert W_{\overrightarrow{O}} \lvert
\end{align}
For a single edged query session, \ie if $\lvert O \lvert = 1$, the support of the corresponding
prefix path $\mathrm{count}(\mathrm{prefix}(\overrightarrow{O})) = \lvert W \lvert$.

Given the query session $Q$, we define $\mathcal{Q} \subseteq \overrightarrow{\mathcal{P}(T)'}$,
the set of all decision paths that can be formed using subsets of edges in $Q$, whose support is
no more than a threshold $\tau$. More formally, \vspace{-1mm}
\begin{align}
\label{eq:allRandPaths}
\hspace{-2mm}\mathcal{Q} = \{\overrightarrow{Q_i} \lvert Q_i \subseteq Q, \mathrm{count}(\overrightarrow{Q_i}) \leq \tau, \mathrm{count}(\mathrm{prefix}(\overrightarrow{Q_i})) > \tau \}
\end{align}

We propose to build a random set of decision paths $\Re \subseteq \mathcal{Q}$, such that
$\lvert \Re \lvert = N$, consisting of only decision paths that are based on the current
query session $Q$, and whose support is no more than $\tau$.
A random decision path $\overrightarrow{Q_i}$ is grown using edges in $Q$ until either
$\mathrm{count}(\overrightarrow{Q_i}) \leq \tau$, or all the edges in $Q$ are exhausted,
whichever comes first. Note that in case all edges
in $Q$ are exhausted before we obtain a path $\overrightarrow{Q_i} \in \mathcal{Q}$, then
$\mathcal{Q} = \phi$. The final score of an edge $e \in C$ for query session $Q$ is given by \vspace{-1mm}
\begin{align}
\label{eq:totalScore}
\mathrm{score}(e) = \frac{1}{\lvert \Re \lvert} \times \sum_{\overrightarrow{Q_i} \in \Re}  \mathrm{supp}(e, Q_i, W)
\end{align}

\begin{algorithm}[t]
\caption{Random Decision Paths Based Edge Suggestion}\label{alg:decisionPaths}
\LinesNumbered
\small

\SetKw{KwAnd}{and}
\SetKw{KwDownTo}{downto}

\KwIn{Data graph $G_d$, Query Log $W$, candidate edges $C$, query session $Q$, number of random decision paths $N$, query log subset threshold $\tau$}

\KwOut{Ranked list of candidate edges}

\BlankLine

$E_{sugg} \leftarrow \phi$, $i \leftarrow 0$;

\While{$i < N$}
{ \label{line:decisionPaths-instantiate-start}
			$\overrightarrow{Q_i} \leftarrow \phi$;
			
			$s_i \leftarrow 0$;
			
			$W_{\overrightarrow{Q_i}} \leftarrow W$;
			
			\While{$s_i < \lvert Q \lvert$}
			{ \label{line:decisionPaths-path-start}
				$e_{rand} \leftarrow \text{sample\_without\_replacement}(Q)$; \label{line:decisionPaths-randsplit}

				$\overrightarrow{Q_i} \leftarrow \overrightarrow{Q_i} \cup \{e_{rand}\}$;
				
				\ForEach{$w \in W_{\overrightarrow{Q_i}}$}
				{
					\If{$e_{rand} \notin w$}
					{
						$W_{\overrightarrow{Q_i}} \leftarrow W_{\overrightarrow{Q_i}} \setminus \{w\}$;
					}
				} \label{line:decisionPaths-wattr}
				
				\If{$\lvert W_{\overrightarrow{Q_i}} \lvert \leq \tau$}
				{
					break;
				}
				
				$s_i \leftarrow s_i + 1$;
			} \label{line:decisionPaths-path-end}
		\ForEach{$e \in C$}
		{ \label{line:decisionPaths-suppcnt-start}
			$\mathrm{supp}(e, Q_i, W) \leftarrow$ Equation~\ref{eq:support};
		
			$E_{sugg} \leftarrow E_{sugg} \cup \{(e, \mathrm{supp}(e, Q_i, W))\}$;
		} \label{line:decisionPaths-suppcnt-end}
	
		$i \leftarrow i+1$;
} \label{line:decisionPaths-instantiate-end}

\ForEach{$e \in C$}
{
    $\mathrm{score}(e) \leftarrow$ Equation~\ref{eq:totalScore};
} \label{line:decisionPaths-choosebest}

/* Return candidate edges by decreasing order of $\mathrm{score}(.)$;*/
\end{algorithm}

Algorithm~\ref{alg:decisionPaths} explains the random decision paths based edge ranking algorithm in detail.
Given a set of candidate edges $C$ and a query session
$Q$, we instantiate $N$ random decision paths (line~\ref{line:decisionPaths-instantiate-start}).
The next edge of the path is chosen uniformly at random without replacement from
$Q$ (line~\ref{line:decisionPaths-randsplit}). The new edge chosen in the path
is used to obtain a subset of entries from the query log $W$.
Only those entries in $W$ that contain all the positive and negative edges in the decision path
$\overrightarrow{Q_i}$ are chosen to be present in
$W_{Q_i}$ (line~\ref{line:decisionPaths-path-start}). A decision path $\overrightarrow{Q_i}$ is
grown until $W_{Q_i}$ contains no more than $\tau$ entries in it (or there
are no more edges to be randomly chosen from in $Q$). The support for each candidate edge $e\in C$ is computed
for each decision path (line~\ref{line:decisionPaths-suppcnt-start}).
The support for each candidate edge is averaged across all the decision paths and the
edges are ranked based on the final score obtained using Equation~\ref{eq:totalScore}
(line~\ref{line:decisionPaths-choosebest}).

\begin{figure}[t]
\centering
\includegraphics[width = 0.65\linewidth, keepaspectratio = true]{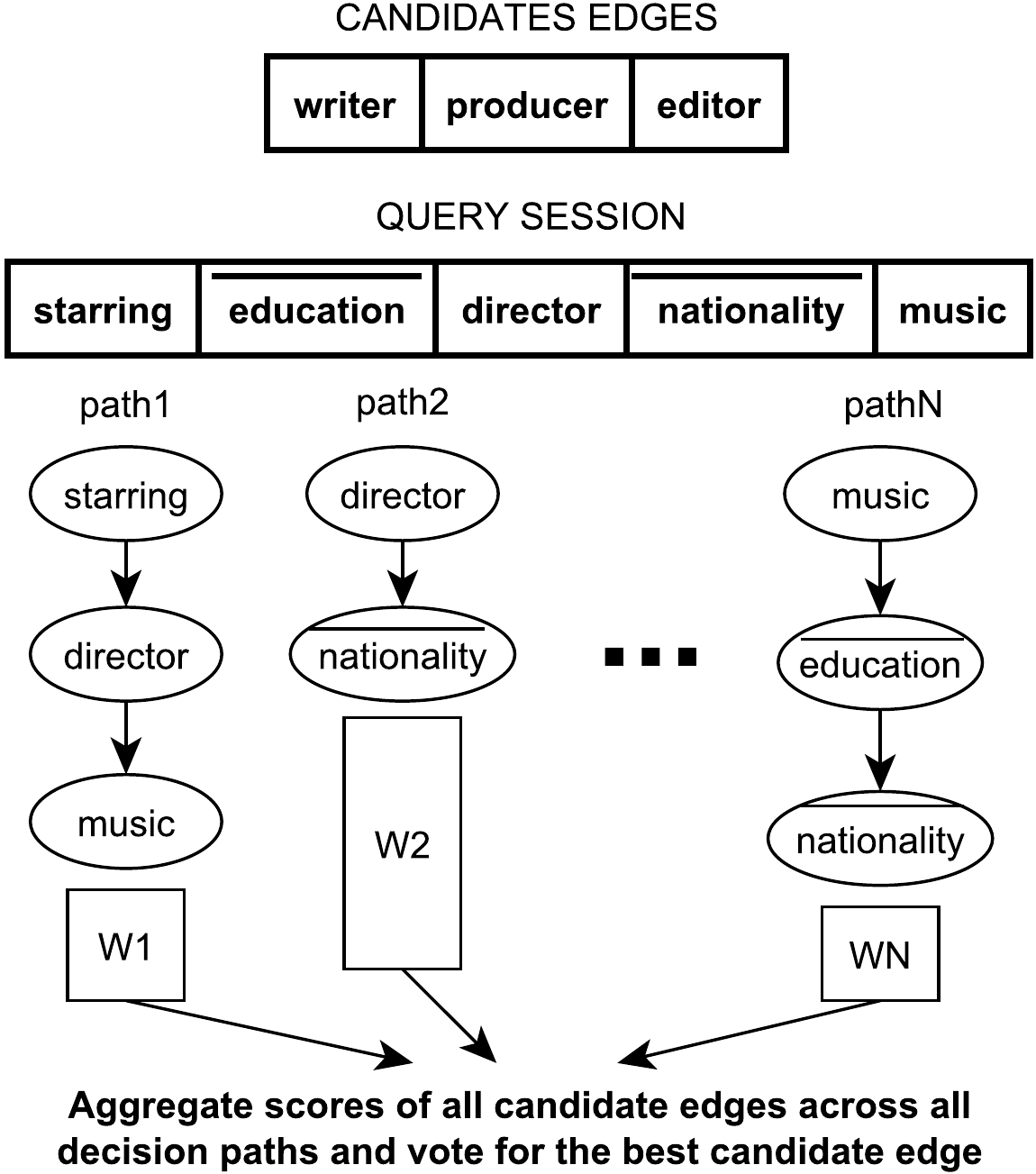}
\vspace{-3mm}
\caption{Random Decision Paths Based Edge Selection}
\label{fig:random-forest}
\end{figure}

Figure \ref{fig:random-forest} shows an example of using random decision paths to rank the candidate edges.
If the set of candidate edges is $C$ = $\{$\edge{writer}, \edge{producer}, \edge{editor}$\}$ and
query session $Q$ contains edges \edge{starring}, $\overline{\edge{education}}$, \edge{director},
$\overline{\edge{nationality}}$, and \edge{music}, $\overrightarrow{path_1}$
through $\overrightarrow{path_N}$ are examples of various random decision paths. For
instance, decision path $\overrightarrow{path_2}$ consists of edges \edge{director} and
$\overline{\edge{nationality}}$, which lead to query log subset $W_{path_2}$ where
$\lvert W_{path_2} \lvert \leq \tau$.
In a decision path $\overrightarrow{path_i}$, the support for each candidate
edge $e \in C$ with entry $\overline{e}$ in $W_{path_i}$ is computed. The support for each candidate
across all the decision paths is aggregated to rank edges in $C$.

\section{Simulating Query Logs}\label{sec:workload}
All the baseline methods and the random decision paths rely on a query log.
But, to the best of our knowledge, a query log for large graphs is not publicly available, except for a SPARQL query log~\cite{dbpedia-sparql}, which is applicable only for the DBpedia data graph.
We thus simulate and bootstrap a query log.
We first find correlated positive edges, using three different methods:
1) using Wikipedia and the data graph, 2) using only the data graph, and 3) using the aforementioned SPARQL query log.
Then negative edges, which indicate edge suggestions that were not accepted by the user, are injected into the simulated query sessions.
If positive edges $e_1$ and $e_2$ are in query session $Q_i$, and another query session $Q_j$ contains $e_1$ but not $e_2$, then $e_2$ is injected into $Q_j$ as a negative edge.

\textbf{Positive edges using Wikipedia and data graph (\system{WikiPos}):}\hspace{2mm}
Each Wikipedia article describes an entity in detail and refers to other Wikiepdia entities by wikilinks.
Given a sentence in a Wikipedia article (or a window of consecutive sentences), the multiple entities mentioned in it can be considered related in some way.
We discover the pairwise relationships between these entities.
Our premise is that these co-occurring relationships simulate the positive edges of a query session.
The intuition is that such consecutive sentences describe closely related facts, and an \system{Orion} user may also have such closely related facts as their query intent.

To find co-occurring positive edges, we map entities mentioned in Wikipedia articles to nodes in the data graph.
Data graphs such as Freebase and DBpedia provide a straight-forward mapping of their nodes to Wikipedia entities.
Given a sentence window, all edges found in the data graph between the mapped entities are approximated to the co-occurring positive edges of a query session in $W$.
We consider all edges between the mapped entities in the data graph, while only a subset of these might actually be mentioned in the corresponding Wikipedia article.
Thus, the co-occurring positive edges identified using this method might be noisy.
We filter out co-occurring positive edges with less support.
Every session in the query log is viewed as an itemset.
We use the Apriori algorithm to generate frequent itemsets, subject to a support $\rho_w$.
The resulting frequent itemsets thus form query sessions with only positive edges.

\textbf{Positive edges using the data graph (\system{DataPos}):}\hspace{2mm}
Another way of finding co-occurring positive edges is to use statistics based on the data graph $G_d$ alone.
For every node $v \in V(G_d)$, an itemset is created which includes all edges incident on $v$ in $G_d$.
This way we converted the graph $G_d$ to $\lvert V(G_d) \lvert$ itemsets.
Here too, we apply the Apriori algorithm to find all frequent itemsets using support $\rho_d$.

\textbf{Positive edges using SPARQL query log (\system{SparqlPos}):}\hspace{2mm}
The DBpedia SPARQL query log~\cite{dbpedia-sparql} contains benchmark queries posed by users on DBpedia through its SPARQL query interface.
We extract co-occurring positive edges using the properties specified in the WHERE clause of the queries.
Since this is a real query log, every set of positive edges found in each WHERE clause is used as is, without applying
any pruning as in \system{WikiPos} and \system{DataPos}.

\textbf{Injecting negative edges to query log (\system{InjectNeg}):}\hspace{2mm}
The aforementioned methods only generate query sessions with positive edges.
But it is crucial to simulate edges that were not accepted by users, since we must rank candidate edges that are correlated with both accepted and ignored edges in a query session.
A simple, but effective strategy is used to introduce negative edges into the query logs.
Consider a query log which has only positive edges, as produced by the aforementioned methods.
For a query session $w \in W$, $T(w)$ is defined as the set of node types of end nodes of all edges in $w$.
I.e., $T(w) = \{t \lvert t\in T_V, \exists e\textsf{=}(u,v) \in E(G_d), \mathrm{etype}(e) \in w \textrm{ s.t. } t \in \mathrm{vtype}(u) \textrm{ or } t \in \mathrm{vtype}(v)\}$.
The set of negative edges added to $w$, denoted $\overline{w}$, is the set of all edges incident on the node types in $T(w)$.
I.e., $\overline{w} = \{\overline{e} \lvert e\textsf{=}(u,v) \in E(G_d), \mathrm{vtype}(u) \in T(w) \textrm{ or } \mathrm{vtype}(v) \in T(w), \mathrm{etype}(e)\notin w\}$.
The new entry for every $w \in W$ consists of $w \cup \overline{w}$, which is then used as the final query log by the various candidate edge ranking methods in Section~\ref{sec:ranking}.

\section{Experiments}\label{sec:experiments}
\vspace{-1mm}\subsection{Setup}
We conducted user studies on a double quad-core 24 GB memory 2.0 GHz Xeon server.
Furthermore, RDP was compared with other edge ranking algorithms (RF, NB, CAR and SVD) on the Lonestar Linux cluster of TACC,~\footnote{\url{https://portal.tacc.utexas.edu/user-guides/lonestar}.} which consists of five Dell PowerEdge R910 server nodes, with four Intel Xeon E7540 2.0GHz 6-core processors on each node, and a total of 1TB memory.

\begin{table}[t]
\centering
\scriptsize
\begin{tabular}{|l|l|l|l|l|}
\hline
\multirow{2}{*}{{\bf Query Log}} & \multicolumn{4}{l|}{{\bf Components Used in Query Log Simulation}} \\ \cline{2-5}
                 &   {\bf Freebase}    &  {\bf DBpedia}    &   {\bf Wikipedia}  & {\bf SPARQL}~\cite{dbpedia-sparql}  \\ \hline
                {\bf Wiki-FB}  &   Yes    &  -     &   Yes  & -  \\ \hline
                {\bf Data-FB}  &   Yes    &  -     &   -   & - \\ \hline
                {\bf Wiki-DB}  &    -  &   Yes    &   Yes  & -  \\ \hline
                {\bf Data-DB}  &    -   &   Yes    &   -  & -  \\ \hline
                {\bf QLog-DB}  &   -    &   -    &    -  & Yes \\ \hline
\end{tabular}
\caption{Query Logs Simulated}
\label{tab:querylogs}
\end{table}

\begin{table}[t]
\centering
\scriptsize
\begin{tabular}{|l|l|}
\hline
{\bf Query Type} & {\bf Query Task} \\ \hline
{\bf Easy} & \parbox{50mm}{Find all Basketball players in Chicago Bulls.} \\ \hline
{\bf Medium} & \parbox{50mm}{Find all award winning films directed by Steven Spielberg.} \\ \hline
{\bf Hard} & \parbox{50mm}{Find all film-actor pairs such that the actor was born in Israel and studied in Harvard University.} \\ \hline
\end{tabular}
\caption{Sample Query Tasks From User Studies}
\label{tab:samplequeries}
\end{table}

\begin{figure}[t]
\centering
  \includegraphics[width = 0.99\linewidth, keepaspectratio = true, scale=0.4]{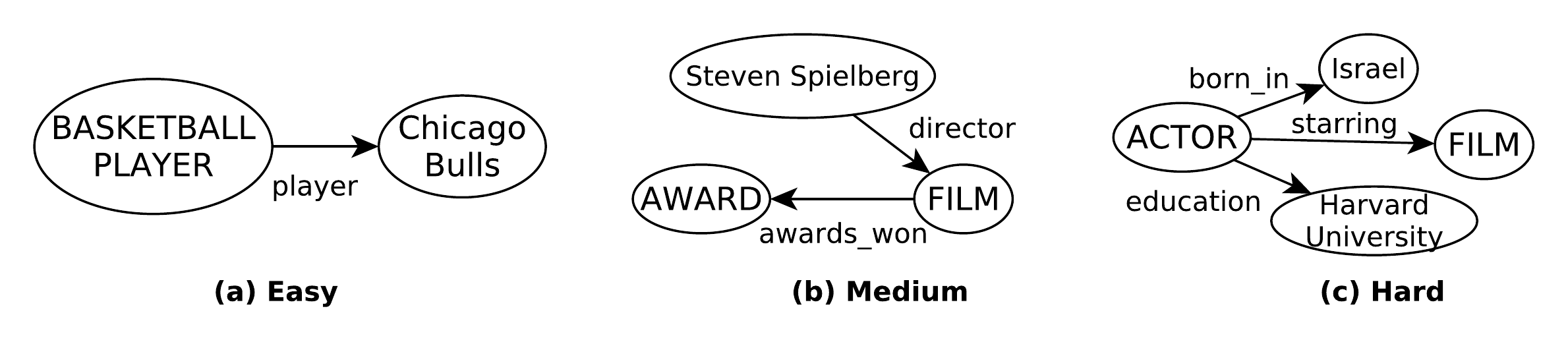}
  \vspace{-2mm}
\caption{Target Query Graphs of Tasks in Table~\ref{tab:samplequeries}}
\label{fig:targetquery}
\end{figure}

\begin{table*}[t]
\centering
\scriptsize
\begin{tabular}{c|l|l|l|l|}
\hline
\multicolumn{1}{|l|}{\parbox{13mm}{{\bf Likert Scale Score}}} &   \parbox{35mm}{{\bf Q1:} How well do you think the query graph formulated by you captures the required query intent?}   &   \parbox{34mm}{{\bf Q2:} How easy was it to use the interface for formulating this query?}  &   \parbox{34mm}{{\bf Q3:} How satisfactory was the overall experience?}  &  \parbox{34mm}{{\bf Q4:} The interface provided features necessary for easily formulating query graphs.}  \\ \hline
\multicolumn{1}{|l|}{1} &  Very Poorly   &   Very Hard  &  Unacceptable   &  Strongly Disagree   \\
\multicolumn{1}{|l|}{2} &   Poorly  &  Hard   &  Poor   &  Disagree   \\
\multicolumn{1}{|l|}{3} &   Adequately  &  Neither Easy Nor Hard   &  Satisfactory   &  Uncertain   \\
\multicolumn{1}{|l|}{4} &   Well  &   Easy  &  Good   &  Agree   \\
\multicolumn{1}{|l|}{5} &   Very Well  &   Very Easy  &  Excellent   &  Strongly Agree   \\ \hline
\end{tabular}
\caption{Survey Questions and Options}
\label{table:survey}
\end{table*}

\textbf{Datasets:}\hspace{2mm}
We used two large real-world data graphs: the 2011 version of Freebase~\cite{Bollacker+08freebase}, and the 2015 version of DBpedia~\cite{AuerBK+07}.
We pre-processed the graphs to keep only nodes that are named entities (\eg \entity{Brad Pitt}), while pruning out nodes corresponding to constant values such as integers and strings among others.
In the original Freebase dataset, every relationship has an inverse relationship in the opposite direction.
For instance, the relationship \edge{director} has \edge{directed by} in the opposite direction.
All such edges in the opposite direction were deleted, since they are redundant.
The resulting Freebase graph contains 30 million nodes, 33 million edges, and 5253 edge types.
After similar pre-processing, the DBpedia graph obtained contains 4 million nodes,
12 million edges and 647 edge types.

\textbf{Query Logs:}\hspace{2mm}
Table~\ref{tab:querylogs} lists the various query logs simulated using the techniques described in Section~\ref{sec:workload}.
One can find positive edges of a query session using different methods, and inject negative edges into them using the method \system{InjectNeg} in Section~\ref{sec:workload}.
We simulated two different query logs for Freebase: Wiki-FB and Data-FB.
The positive edges for Wiki-FB were simulated using both Wikipedia (September 2014 version) and the Freebase data graph, and the positive edges for Data-DB were simulated using only the Freebase data graph, by methods \system{WikiPos} and \system{DataPos} in Section~\ref{sec:workload}, respectively.
We simulated three different query logs for DBpedia: Wiki-DB, Data-DB and
QLog-DB.
Wiki-DB and Data-DB were simulated via the same approach for Wiki-FB and Data-FB, except that DBpedia (instead of Freebase) was the data graph.
For QLog-DB, the positive edges were simulated by \system{SparqlPos} in Section ~\ref{sec:workload}.

\textbf{Systems Compared in User Studies:}\hspace{2mm}
To verify if \system{Orion} indeed makes it easier
for users to formulate query graphs, we conducted user studies with two different user interfaces: \system{Orion}, and \system{Naive}.
\system{Orion} operates in both passive and active modes (cf. Section~\ref{sec:ui-overview}).
\system{Naive} on the other hand does not make any automatic suggestions and only lets users manually add nodes and edges on the canvas.
The various candidate edges are sorted alphabetically and presented to the user in a drop down list.
This mimics the query formulation support offered in existing visual query systems such as~\cite{quble}.

\textbf{Methods Compared for Ranking Candidate Edges:}\hspace{2mm}
We compared the effectiveness of \system{Orion}'s candidate edge ranking algorithm (RDP) with the baseline methods described in Section~\ref{sec:baselineMethods}, including RF, NB, CAR and SVD.

\subsection{User Studies}\label{sec:exp-userstudies}
\textbf{User Study Set-up:}\hspace{2mm}
We conducted an extensive user study with 30 graduate students in the authors' institution.
The students neither had any expertise with graph query formulation, nor did they have exposure to the data graphs.
None of these students were exposed to this research in any way other than participating in the user study.
We conducted A/B testing using the two interfaces, \system{Orion} and \system{Naive}.
The underlying data graph for both systems was Freebase, and were hosted online on the aforementioned Xeon server.
We arbitrarily chose 15 students to work with \system{Orion}, and the other 15 students worked with \system{Naive}.
The users of \system{Orion} were not exposed to \system{Naive}, and vice versa.
We created a pool of 21 query tasks, which consisted of three levels of difficulty.
9 queries were \emph{easy}, 6 queries were \emph{medium} and 6 queries were \emph{hard}.
The target query graphs for each easy and medium query tasks had exactly one and two edges, respectively.
The target query graphs for hard query tasks had at least three and at most 5 edges.
Table~\ref{tab:samplequeries} lists one sample query for each of the three categories.
Figures~\ref{fig:targetquery}(a), (b) and (c) depict the target query graphs for the query tasks listed in Table~\ref{tab:samplequeries}.

We created 15 different query sheets, where each consisted of 3 easy, 2 medium and 2 hard query tasks,
chosen from the pool of 21 queries designed. Each \system{Orion} and \system{Naive} user
was given a query sheet as the task set to complete which ensured that users of both
systems worked on the same query tasks. Each user was given an initial 15-minute introduction by
the moderators regarding the data graphs, graph query formulation, and the user interface.
The users then spent 45 minutes working on their respective query sheets.
The users were allowed to ask any clarification questions regarding the tasks during the user study.
Each user was awarded a gift card worth $\$15.00$ for their participation in the user study.
Since 15 users worked on 7 queries each, we obtained a total of 105 responses for both \system{Orion} and \system{Naive}.

\textbf{Survey Form:}\hspace{2mm}
The users were requested to fill an online survey form at the end of each query task, thus resulting in
105 different survey form responses for each user interface. The survey form had four questions:
$Q1$, $Q2$, $Q3$ and $Q4$, as listed in Table~\ref{table:survey}. Each question had five options,
specifying the level of agreement a user
could have with the particular aspect of the interface measured by the question. We assign a score for
every option in each question based on the Likert scale shown in Table~\ref{table:survey}.
The least favourable experience with respect to each question is assigned a score of 1, and the most
favoured experience is assigned a score of 5.

\begin{table}[t]
\centering
\scriptsize
\begin{tabular}{|l|l|l|l|l|l|}
\hline
{\bf System} & {\bf Queries} & \parbox{11mm}{{\bf Sample Size}} & \parbox{14mm}{{\bf Conversion Rate} ($c$)} &  {\bf z-value}  &  {\bf p-value}  \\ \hline
\system{Orion} & \multirow{2}{*}{All} & \multirow{2}{*}{105} & $c_O$=0.74 & \multirow{2}{*}{0.92} & \multirow{2}{*}{0.1788} \\ \cline{1-1} \cline{4-4}
\system{Naive} &          &         & $c_N$=0.68 &                   &                   \\ \hline
\system{Orion} & \multirow{2}{*}{\parbox{11mm}{Medium + Hard}} & \multirow{2}{*}{60} & $c_O$=0.70 & \multirow{2}{*}{{\bf 1.36}} & \multirow{2}{*}{{\bf 0.0869}} \\ \cline{1-1} \cline{4-4}
\system{Naive} &          &         & $c_N$=0.58 &                   &                   \\ \hline
\end{tabular}
\caption{Conversion Rates of \system{Naive} and \system{Orion}}
\label{tab:userstudy-conversion}
\end{table}

\subsubsection{Efficiency Based on Conversion Rate}
\textbf{Measure:}\hspace{2mm}
One of the popular metrics used to measure the effectiveness of the systems compared in A/B testing
is conversion rate $c$, which is the percentage of tasks completed
successfully by users. The conversion rate is defined over a set of $\mathrm{Tasks}$ as:
\begin{align}
c = \frac{\sum_{\mathrm{task} \in \mathrm{Tasks}}{\mathrm{sim}(G_u,G_t)}}{\lvert \mathrm{Tasks} \lvert}
\end{align}
where $\mathrm{task}$ is a query task assigned to the user, $G_u$ is the corresponding query graph constructed by
the user, and $G_t$ is the actual target query graph corresponding to $\mathrm{task}$. The similarity measure
$\mathrm{sim}(G_u,G_t)$ captures the notion of success, based on how similar $G_u$ is to $G_t$.
Since we designed the query tasks, the target query graph for each query task was known to us apriori.
The query graph constructed by each user was recorded by the interface during the user
study. Intuitively, the similarity between $G_u$ and $G_t$ is based on the edge-preserving sub-graph
isomorphic match between the two graphs. More formally, $sim(G_u,G_t)$ is defined as: \vspace{-1mm}
\begin{align}
\label{eq:ranking_function}
\mathrm{sim}(G_u, G_t) = \frac{\max_{f}{\sum_{\substack{ e=(u,v) \in E(G_u) \\ e'=(f(u), f(v)) \in E(G_t)}}} \mathrm{match}(e, e')}{\lvert E(G_t) \lvert}
\end{align}
where $f:V(G_u) \rightarrow V(G_t)$ is a bijection, and $\mathrm{match}(e, e')$ is a matching function defined as: \vspace{-1mm}
\begin{multline}
\label{eq:match}
\hspace{-4mm} \mathrm{match}(e,e')\text{=}
	\begin{cases}
	1 & \text{if } u\text{=}f(u), v\text{=}f(v), etype(e)=etype(e')\\
	0 & \text{otherwise}
	\end{cases}\vspace{-2mm}
\end{multline}

\textbf{Results:}\hspace{2mm}
Table~\ref{tab:userstudy-conversion} summarizes the conversion rates of \system{Orion} and
\system{Naive} over the set of all query tasks (easy, medium and hard query tasks), and also over only
the medium and hard query tasks. We observe that \system{Orion}
has a better conversion rate than \system{Naive} in both scenarios. But, on performing a two sample
Z-test with significance level $\alpha$=0.1, only the observation that \system{Orion} has a better
conversion rate than \system{Naive} for medium and hard queries is statistically significant.
We next describe the hypothesis testing of the two scenarios in detail.

\begin{figure*}[t]
\begin{minipage}[b]{0.70\linewidth}
\centering
\subfigure[All queries]{
\includegraphics[scale=0.17, angle=360]{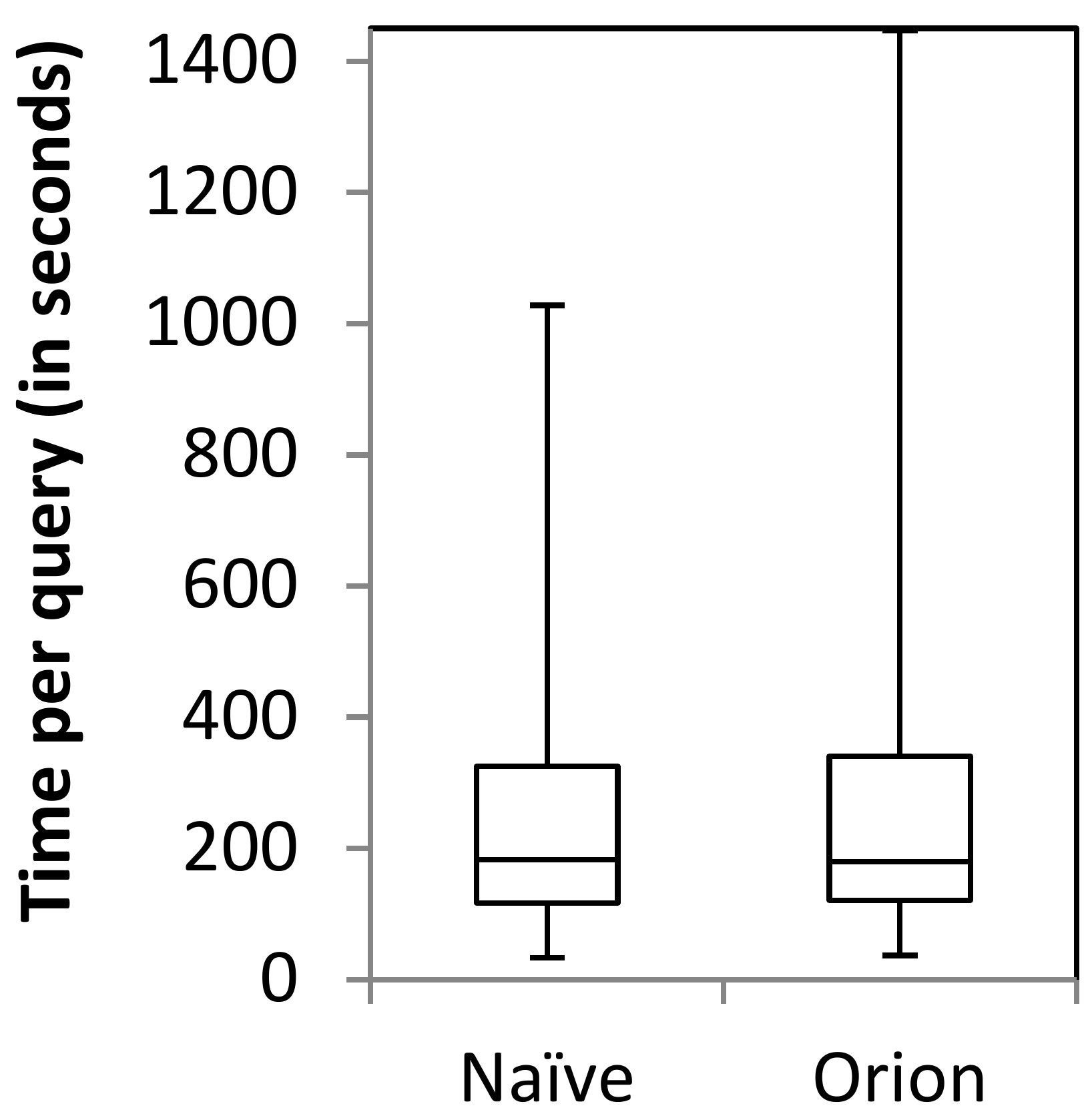}\vspace{1mm}
\label{fig:userstudy-time-all}
}
\subfigure[Easy queries]{
\includegraphics[scale=0.17, angle=360]{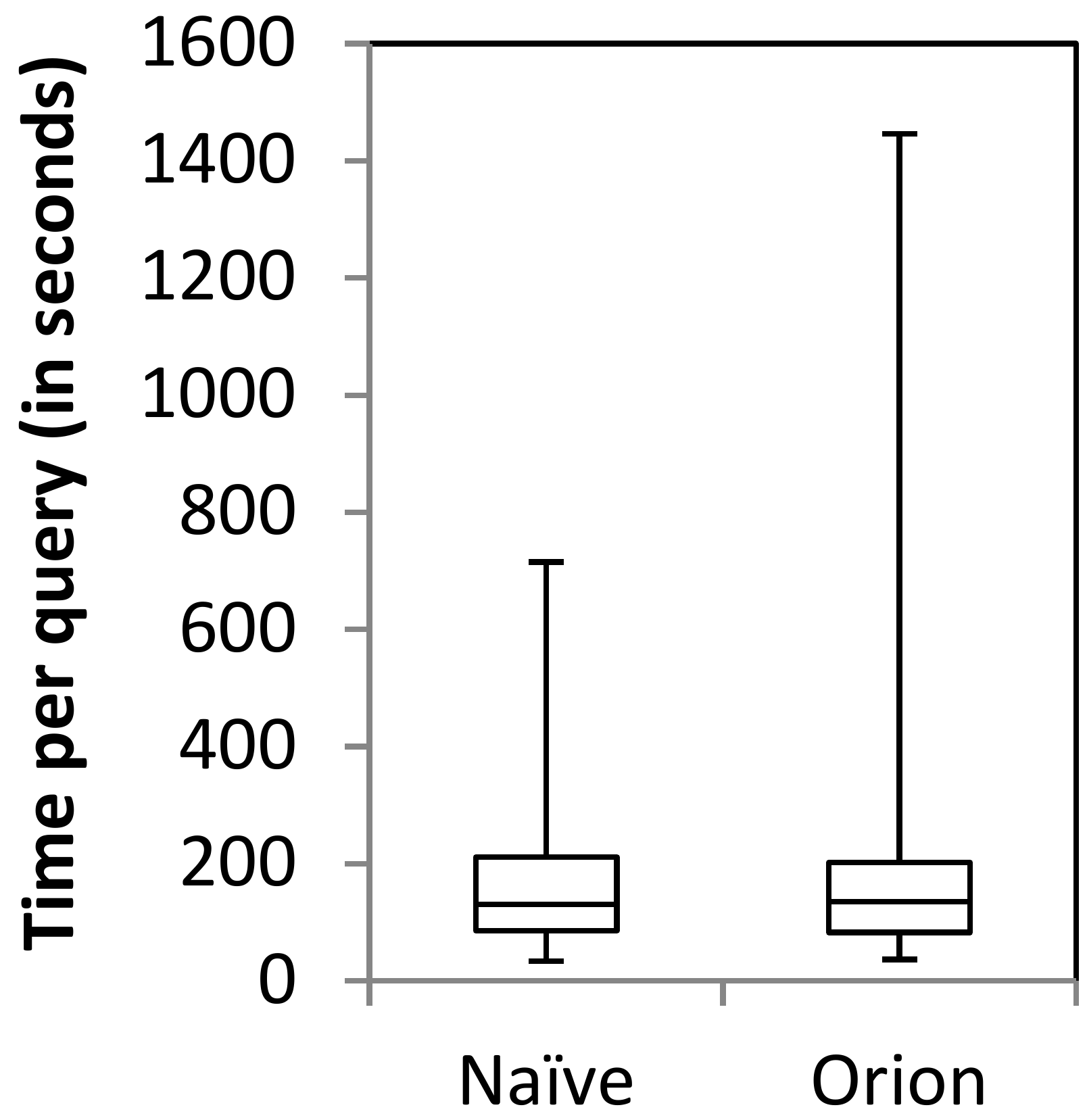}\vspace{1mm}
\label{fig:userstudy-time-easy}
}
\subfigure[Medium queries]{
\includegraphics[scale=0.17, angle=360]{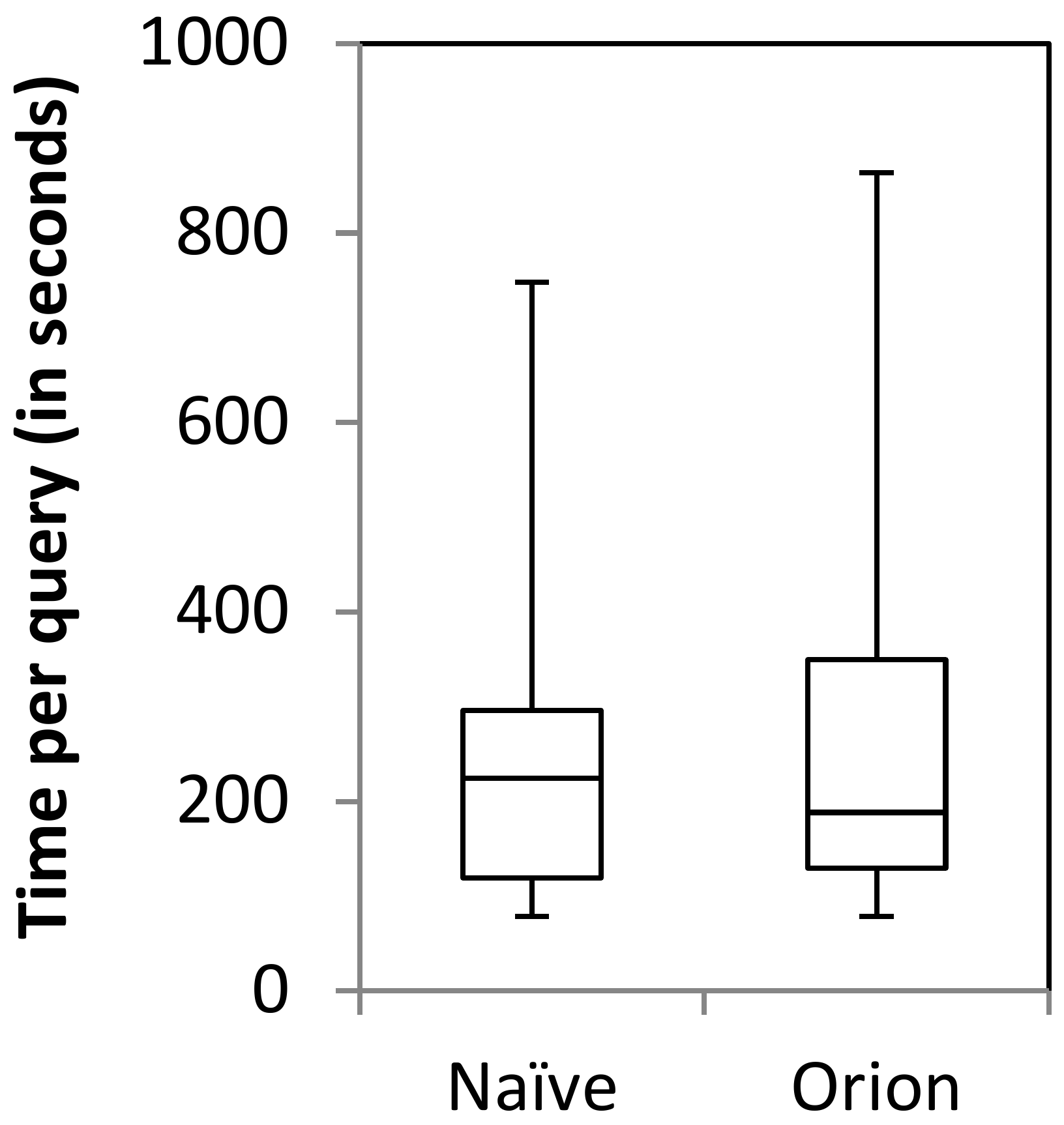}\vspace{1mm}
\label{fig:userstudy-time-med}
}
\subfigure[Hard queries]{
\includegraphics[scale=0.17, angle=360]{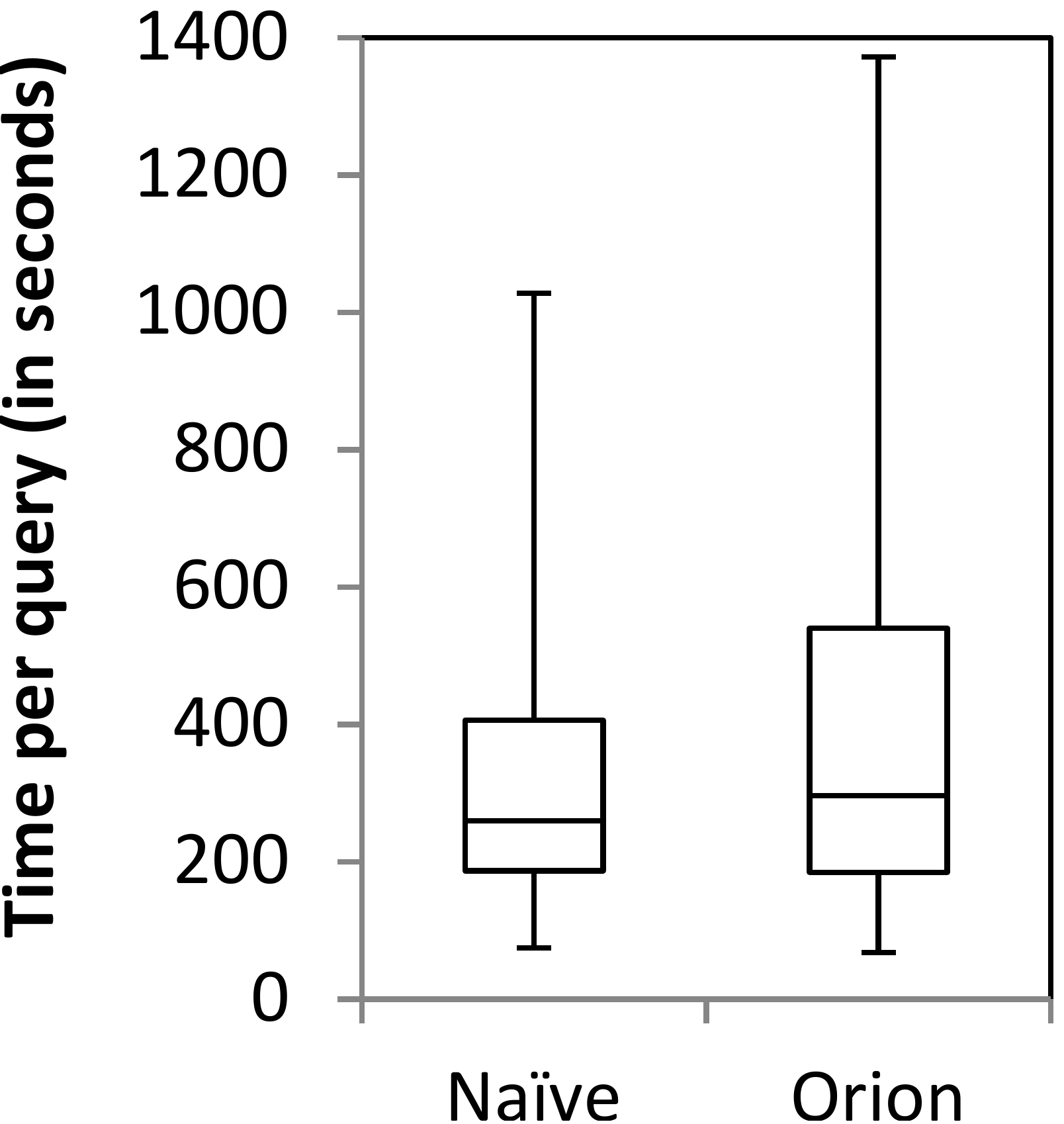}\vspace{1mm}
\label{fig:userstudy-time-hard}
}
\caption{User Studies Efficiency Based on Time: \system{Naive} and \system{Orion}}
\label{fig:userstudy-time}
\end{minipage}\vspace{-3mm}
\begin{minipage}[b]{0.30\linewidth}
\centering
  \includegraphics[scale=0.30, angle=360]{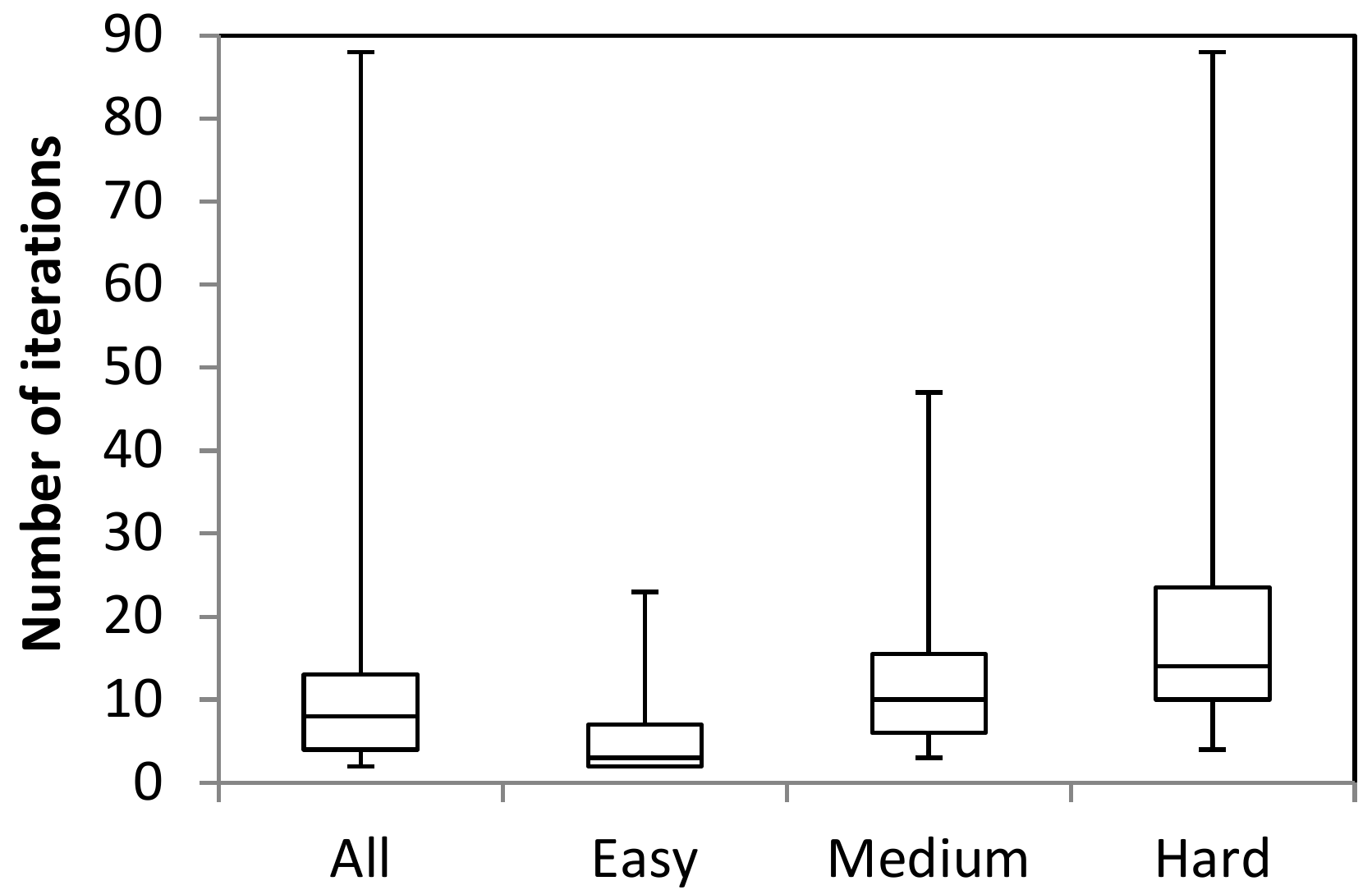}\vspace{-2mm}
\caption{User Studies Efficiency Based on Iterations: \system{Orion}}
\label{fig:userstudy-iters}
\end{minipage}\vspace{-3mm}
\end{figure*}

The conversion rate of \system{Orion}, $c_O$, over all the 105 query tasks is 0.74, and the conversion
rate of \system{Naive}, $c_N$, for the same set of tasks is 0.68. On average, \system{Orion} users had
a higher chance of formulating the correct query graph compared to the \system{Naive} users.
We assume that constructing a query graph follows a Bernoulli trial, with the
probability of successfully constructing the target query graph on \system{Orion} and \system{Naive}
as $p_O = c_O$ and $p_N = c_N$ respectively.
Our hypothesis, $H_{A1}$, is that \system{Orion} has a better conversion rate than \system{Naive}: $H_{A1}$: $p_{O} > p_{N}$.
The null hypothesis $H_{01}$ is given by $H_{01}$: $p_{O} \leq p_{N}$.
For the aforementioned conversion rates of \system{Orion} and \system{Naive},
and a sample size of 105, $z = 0.92$. This results in a p-value of 0.1788. Since the p-value
$ > \alpha$, the null hypothesis cannot be rejected as the data does not significantly support
our hypothesis.

We dive in deeper to investigate if there are scenarios where \system{Orion} does perform better
than \system{Naive}. The conversion rate of only medium and hard query tasks (which is equal to a total
of 60 query tasks) for \system{Orion} is 0.70, and is equal to 0.58 for \system{Naive}, \ie
$c_O = p_O = 0.70$ and $c_N = p_N = 0.58$. This indicates that \system{Orion} users have a better
chance of successfully constructing query graphs with two or more edges, compared to \system{Naive} users.
Our new hypothesis, $H_{A2}$, is that \system{Orion} has a better conversion rate than \system{Naive} for medium and hard queries: $H_{A2}$: $p_{O} > p_{N}$.
The null hypothesis $H_{02}$ is given by $H_{02}$: $p_{O} \leq p_{N}$.
For the aforementioned conversion rates of \system{Orion} and \system{Naive}, and a sample size of 60,
$z = 1.36$, resulting in a p-value of 0.0869. Since the p-value
$ < \alpha$, the data significantly supports our claim that \system{Orion} users have a higher chance
of successfully constructing complex query graphs containing two or more edges.

\subsubsection{Efficiency Based on Time}
We next measure the time taken by a user to construct the query graph for a given query task:
the time elapsed between the first time a user
clicks on the query canvas for a new query task, to the time the user clicks on the "Submit"
button of the interface. This was recorded in the background during the user study.
Figure~\ref{fig:userstudy-time-all} shows the distribution of the time taken to complete a query task.
We observe that half of the 105 query tasks were completed within 180 seconds by \system{Orion} users,
while \system{Naive} users completed the same number of query tasks within 183.2 seconds. Around 26 query
tasks were completed between 180 to 340.5 seconds, and between 183.24 to 325.7 seconds by \system{Orion}
and \system{Naive} users respectively.
Although, there were a few query tasks that took a long time to be completed, with
a maximum of 1446.3 seconds for \system{Orion} users and 1027.8 seconds for \system{Naive} users.
We further study the distribution of the time taken to complete query tasks based on the level of difficulty of
the tasks. Figure~\ref{fig:userstudy-time-easy} compares the time taken for easy query tasks. We observe that
around 23 of the 45 easy queries are completed within 135.5 and 130.3 seconds by \system{Orion} and
\system{Naive} users respectively. Another 12 queries were completed between 135.5 to 202.3 seconds by
\system{Orion} users, and between 130.3 to 211.3 seconds by \system{Naive} users.
Figure~\ref{fig:userstudy-time-med} compares the time taken for medium query tasks. We observe that
around 15 of the 30 medium queries are completed within 188.2 and 224.6 seconds by \system{Orion} and
\system{Naive} users respectively. Another 7 queries were completed between 188.2 to 349.6 seconds by \system{Orion}
users, and between 224.6 to 296.2 seconds by \system{Naive} users. Finally, Figure~\ref{fig:userstudy-time-hard}
compares the time taken for hard query tasks. We observe that
around 15 of the 30 hard queries are completed within 296.1 and 259.6 seconds by \system{Orion} and
\system{Naive} users respectively. Another 7 queries were completed between 296.1 to 540.4 seconds by \system{Orion}
users, and between 259.6 to 406.4 seconds by \system{Naive} users. We observe that despite the steeper learning curve
of \system{Orion} due to the superior number of features in it, the time taken to complete a majority of the query tasks is
comparable with that of \system{Naive}.

\subsubsection{Efficiency Based on Number of Iterations}
We next measure the effectiveness of \system{Orion} using the number of iterations involved in the query construction process: the number of times a ranked list of edges is presented to the user.
The number of iterations is incremented in one of three ways: 1) the user selects one or more of the automatically suggested edges
in active mode, and clicks on the canvas to get the next set of suggestions, 2) the user ignores all the suggestions
made in active mode and clicks on "Refresh Suggestions" to get a new set of automatic suggestions, and 3) the user
draws a new edge in passive mode. We do not measure this for \system{Naive} since there are no automatic
ranked suggestions made in it.
Figure~\ref{fig:userstudy-iters} shows the distribution
of the number of iterations required to construct query graphs. Overall, \system{Orion} users needed no more than only 13
iterations to complete around 79 of the 105 queries. Half of the easy, medium and hard queries required no more than
3, 10 and 14 iterations respectively. Another 11 easy queries required between 3 to 7 iterations, while 7 medium and hard
queries each required between 10 to 15.5 and 14 to 23.5 iterations respectively.
This indicates that the features offered by \system{Orion} helped users formulate query graphs with few interactions with
the interface.

\begin{figure*}[t]
\centering
\subfigure[All queries]{
\includegraphics[scale=0.24, angle=360]{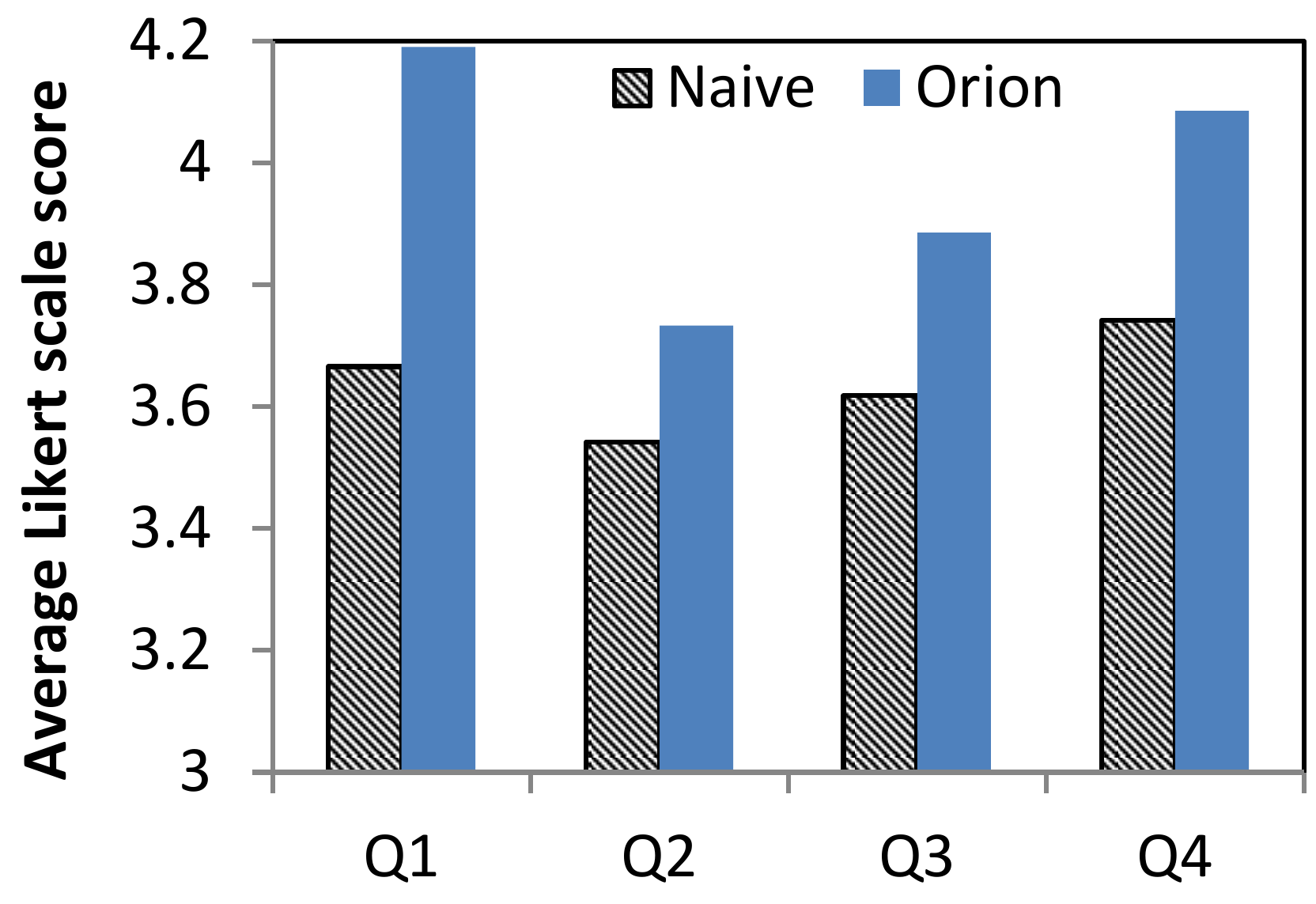}\vspace{1mm}
\label{fig:userstudy-survey-all}
}
\subfigure[Easy queries]{
\includegraphics[scale=0.24, angle=360]{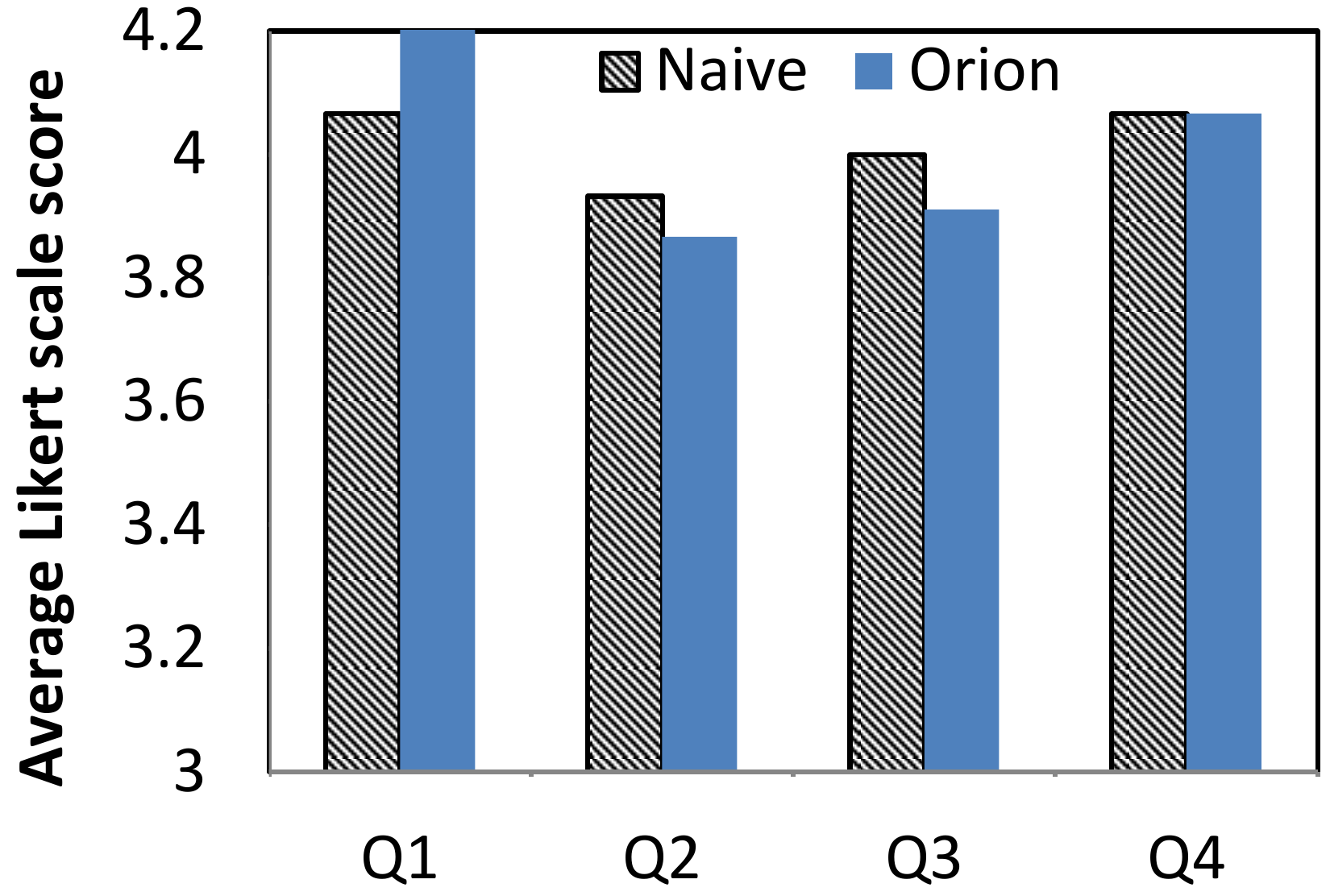}\vspace{1mm}
\label{fig:userstudy-survey-easy}
}
\subfigure[Medium queries]{
\includegraphics[scale=0.24, angle=360]{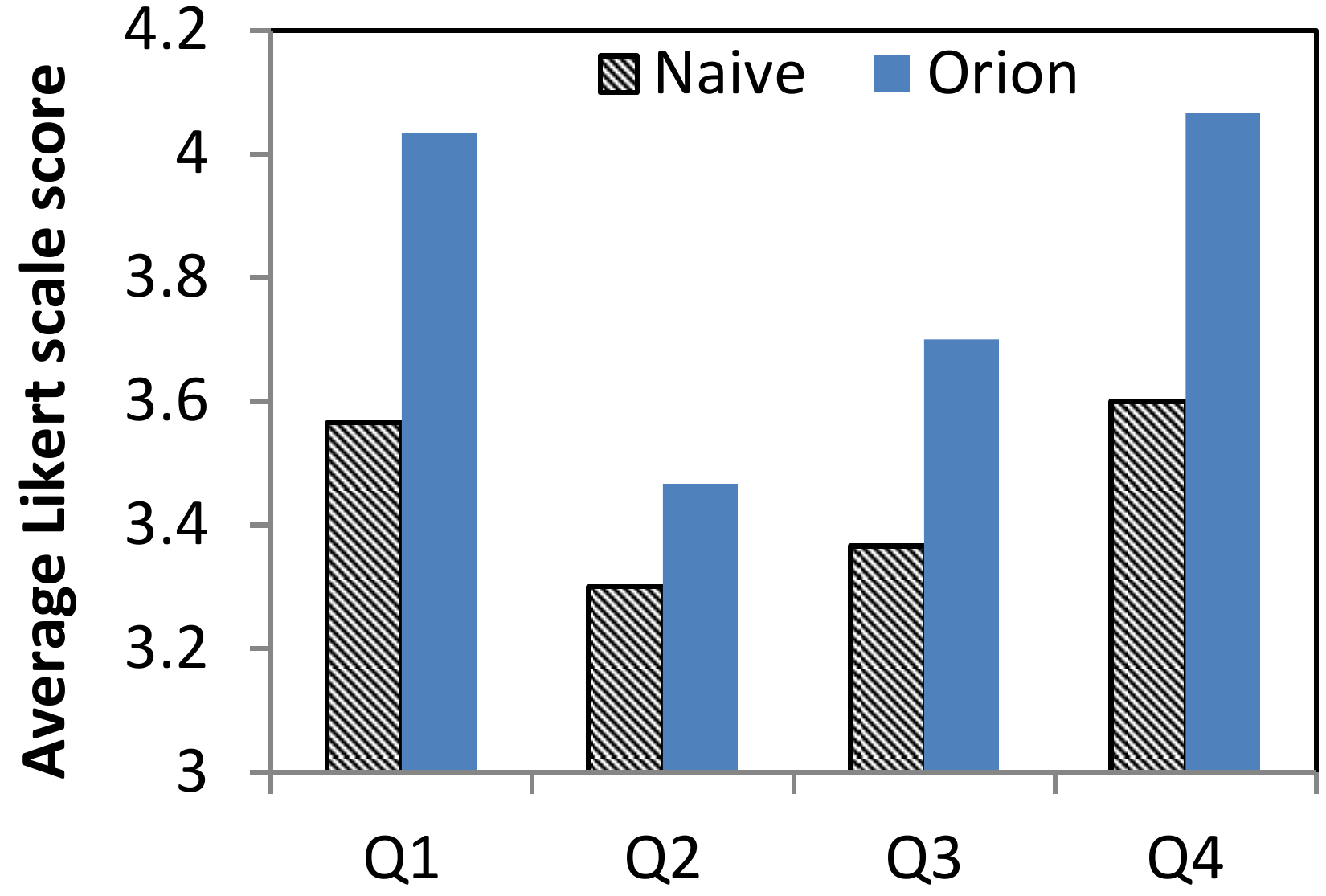}\vspace{1mm}
\label{fig:userstudy-survey-med}
}
\subfigure[Hard queries]{
\includegraphics[scale=0.24, angle=360]{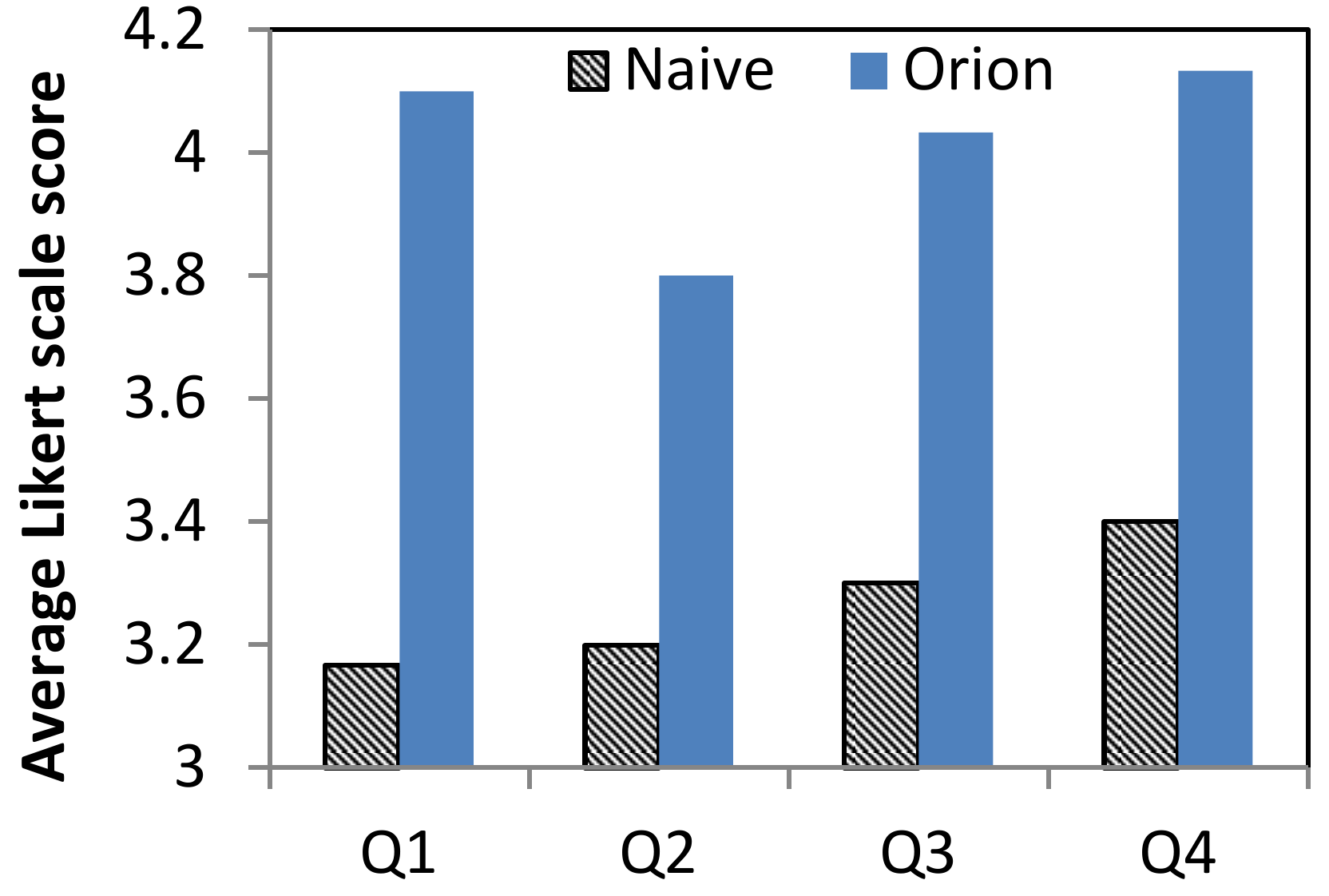}\vspace{1mm}
\label{fig:userstudy-survey-hard}
}
\vspace{-3mm}
\caption{User Experience Based on Survey Responses}
\vspace{-3mm}
\label{fig:userstudy-survey}
\end{figure*}

\subsubsection{User Experience Results}
The user experience results is based on the answers to all the questions in the survey form
by all the users. The overall user experience for each question of an interface is measured
by averaging the score obtained for that question across all the users working on that interface.
Figure~\ref{fig:userstudy-survey-all} shows the overall user response of all the questions, across
all the 105 users for both \system{Orion} and \system{Naive}. We observe that
\system{Orion} users report an improvement of 0.5 for $Q1$, 0.2 for $Q2$, 0.25
for $Q3$ and 0.3 for $Q4$ on Likert scale, when compared to the \system{Naive} users.

We further break down the average score over each question based on the difficulty level of the
query task to study the difference in user experience between \system{Orion} and \system{Naive} in detail.
Figure~\ref{fig:userstudy-survey-easy} shows the average score over only the easy query tasks (a total of 45
query tasks each for both \system{Orion} and \system{Naive}), which shows
that \system{Orion} users had a better experience than the \system{Naive} users w.r.t $Q1$,
while the \system{Naive} users had a slightly better experience than \system{Orion} users w.r.t $Q2$
and $Q3$. Both the sets of users had similar experience w.r.t $Q4$.
Figure~\ref{fig:userstudy-survey-med} shows the average score over only the medium query tasks (a total
of 30 query tasks each for both \system{Orion} and \system{Naive}), which shows
that \system{Orion} users had an improvement of 0.4 on Likert scale w.r.t $Q1$ and $Q4$ compared
to the \system{Naive} users. They also had an improvement close to 0.1 on Likert scale w.r.t
both $Q2$ and $Q3$. Finally, Figure~\ref{fig:userstudy-survey-hard} shows the average score over only
the hard query tasks (a total of 30 query tasks each for both \system{Orion} and \system{Naive}),
which shows that \system{Orion} users felt a significant improvement in the user experience across
all four questions. \system{Orion} users had an improvement of around 1.0 w.r.t $Q1$, 0.6 w.r.t
$Q2$, and 0.7 w.r.t both $Q3$ and $Q4$.
We thus observe that as the difficulty level of the query graph being constructed increases, the
usability of \system{Orion} seems significantly better than \system{Naive}'s. \system{Naive} users
find the system uncomfortable to use when the target query graph contains two or more edges.

\begin{figure*}[t]
\begin{minipage}[b]{0.50\linewidth}
\centering
\subfigure[Freebase]{
\includegraphics[scale=0.2, angle=360]{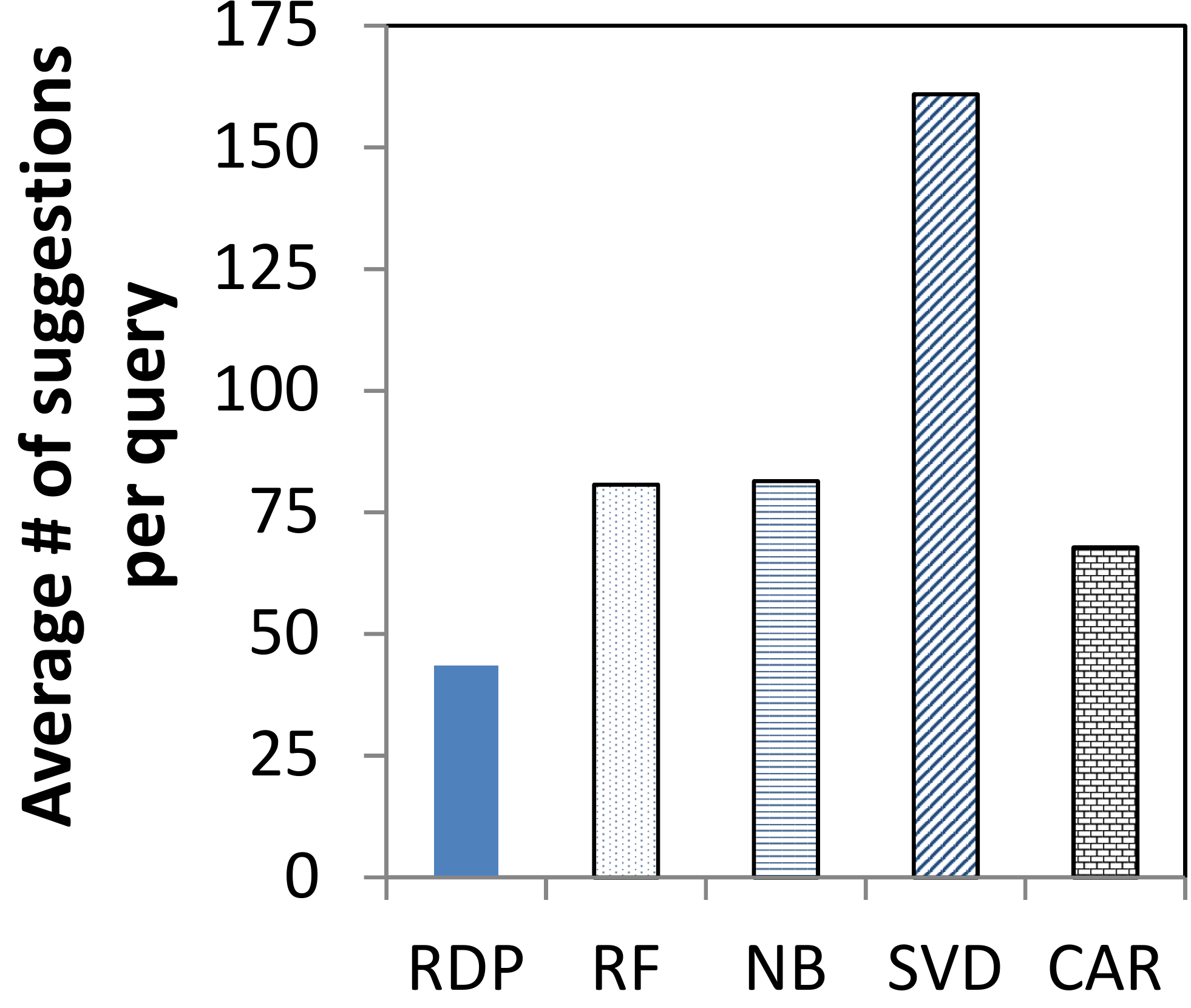}\vspace{1mm}
\label{fig:algo-iters-fb}
}
\hspace{2mm}
\subfigure[DBpedia]{
\includegraphics[scale=0.2, angle=360]{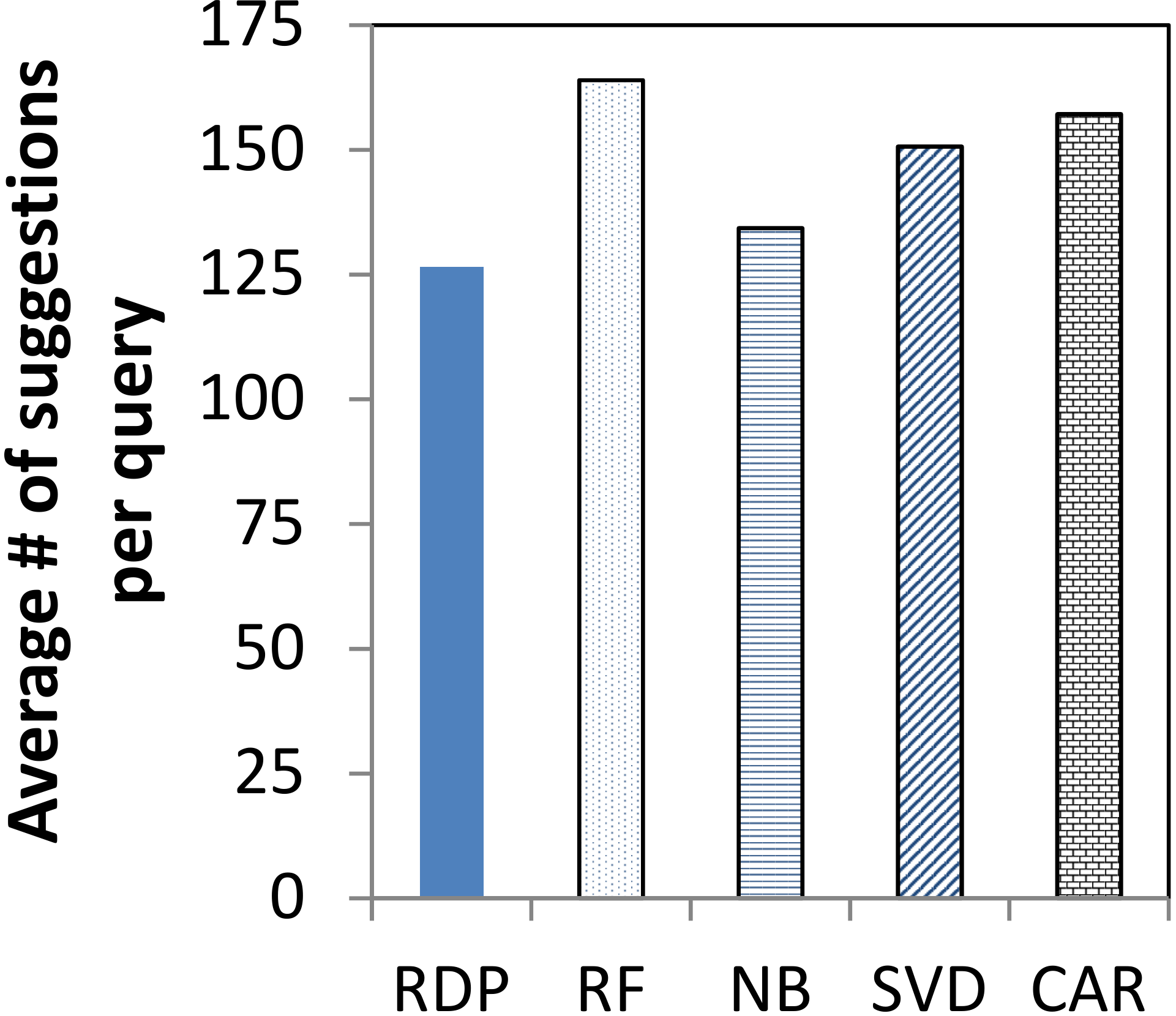}\vspace{1mm}
\label{fig:algo-iters-db}
}
\vspace{-3mm}
\caption{Efficiency of All Methods: Number of Suggestions}
\vspace{-3mm}
\label{fig:algo-iters}
\end{minipage}
\begin{minipage}[b]{0.50\linewidth}
\centering
\subfigure[Freebase]{
\includegraphics[scale=0.22, angle=360]{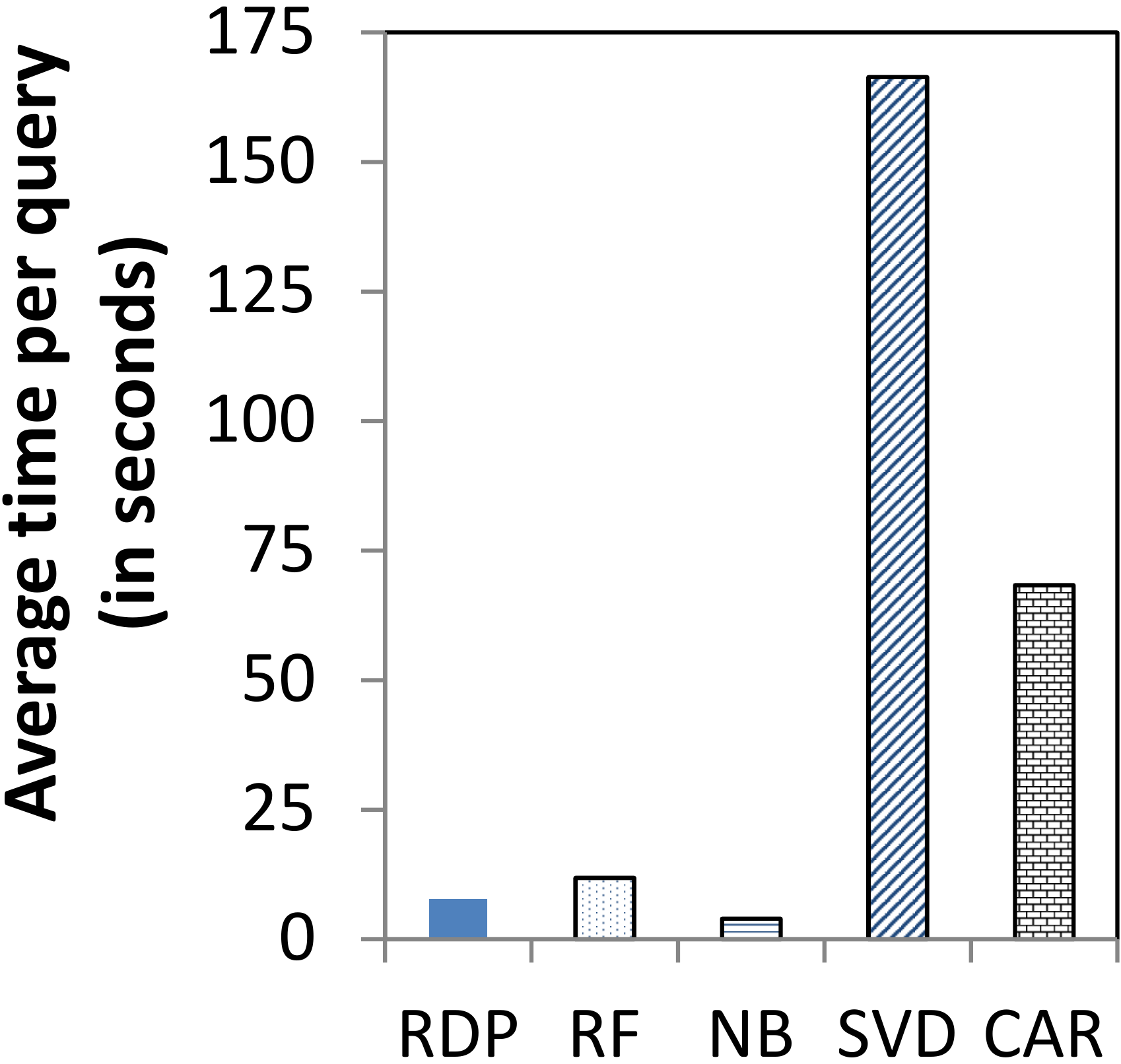}\vspace{1mm}
\label{fig:algo-time-fb}
}
\hspace{1mm}
\subfigure[DBpedia]{
\includegraphics[scale=0.22, angle=360]{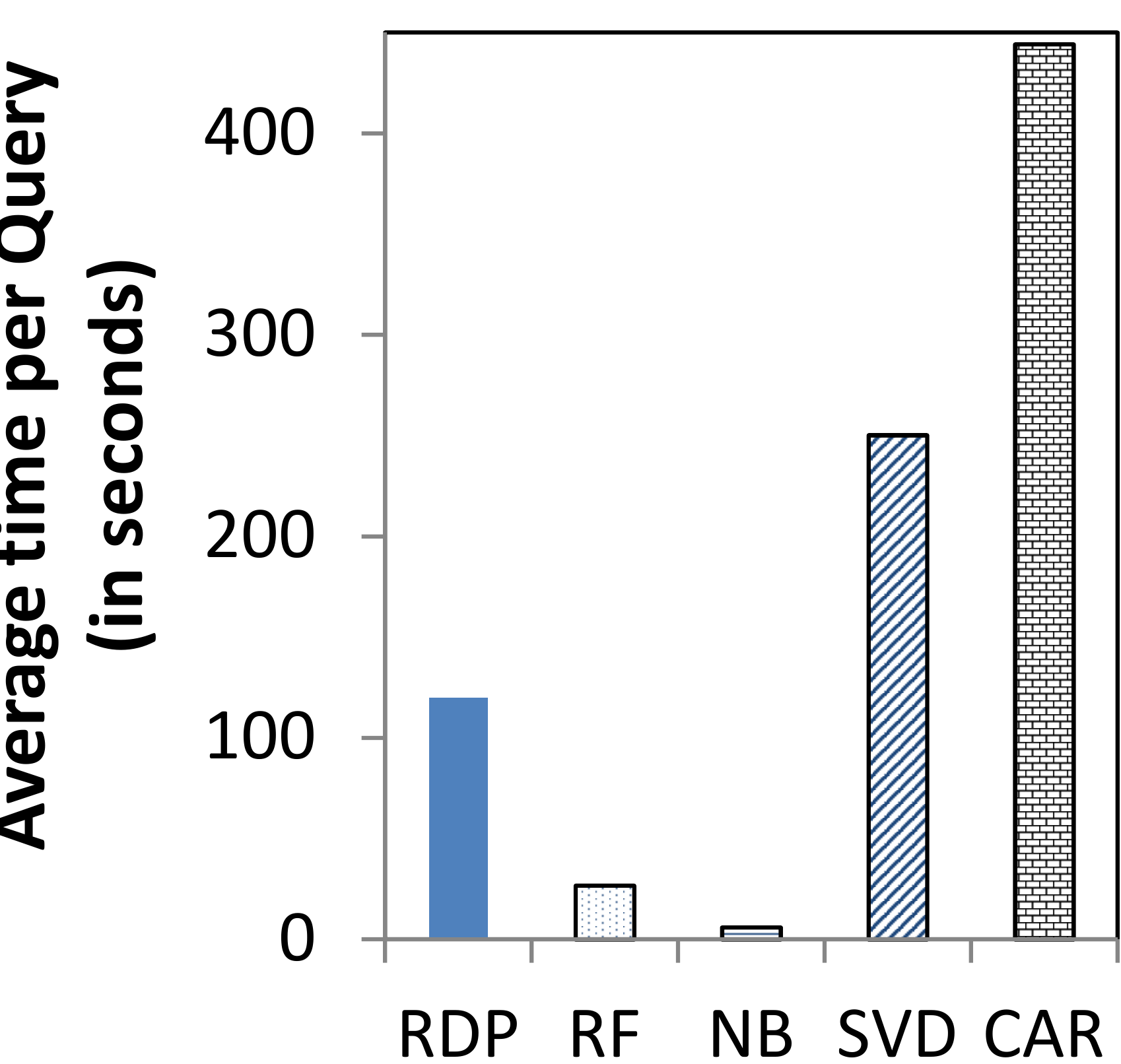}\vspace{1mm}
\label{fig:algo-time-db}
}
\vspace{-3mm}
\caption{Efficiency of All Methods: Time}
\vspace{-3mm}
\label{fig:algo-time}
\end{minipage}
\end{figure*}

\begin{figure*}[t]
\begin{minipage}[b]{0.45\linewidth}
\centering
\subfigure[Freebase]{
\includegraphics[scale=0.18, angle=360]{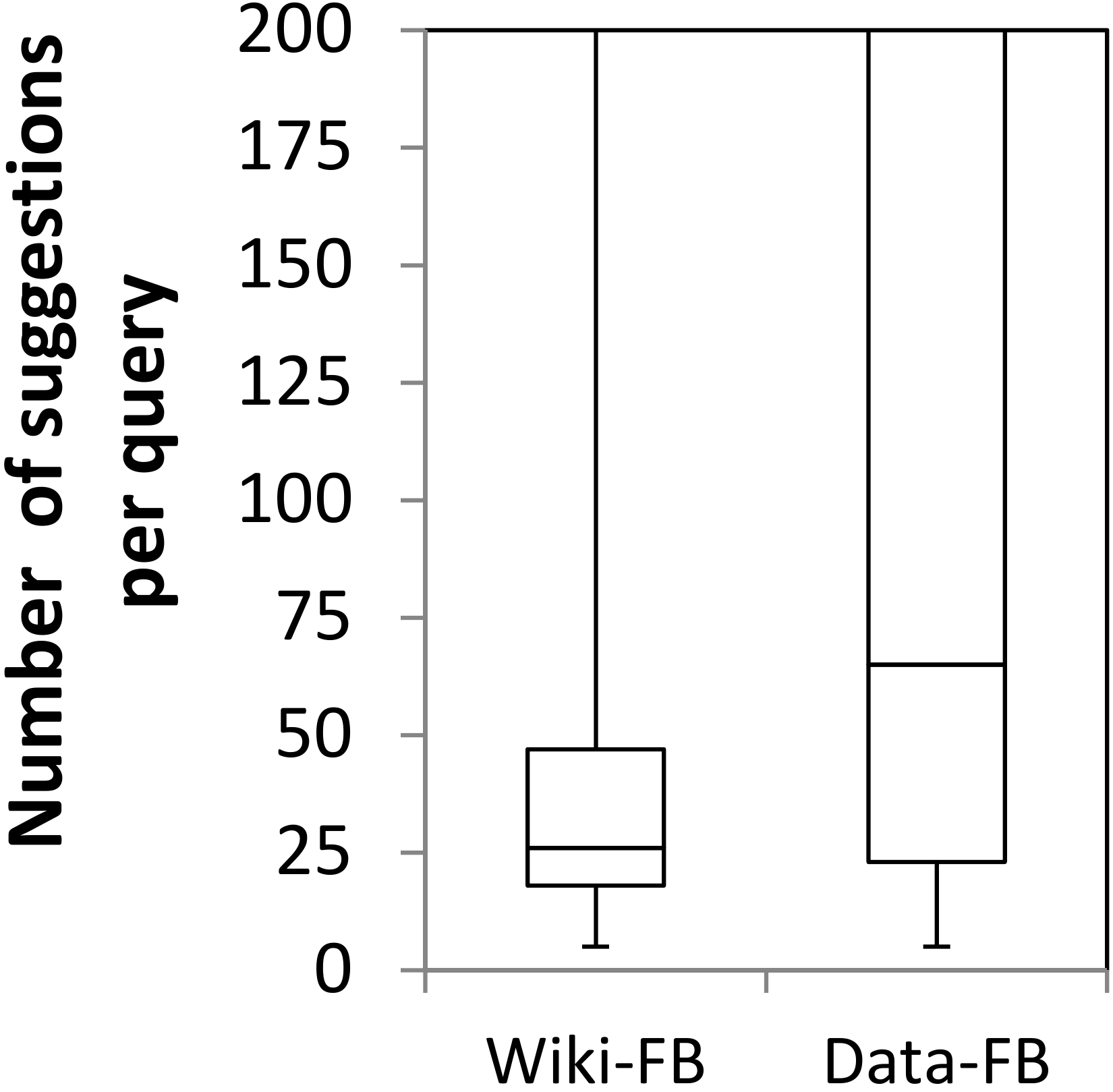}\vspace{1mm}
\label{fig:algo-log-fb}
}
\hspace{1mm}
\subfigure[DBpedia]{
\includegraphics[scale=0.2, angle=360]{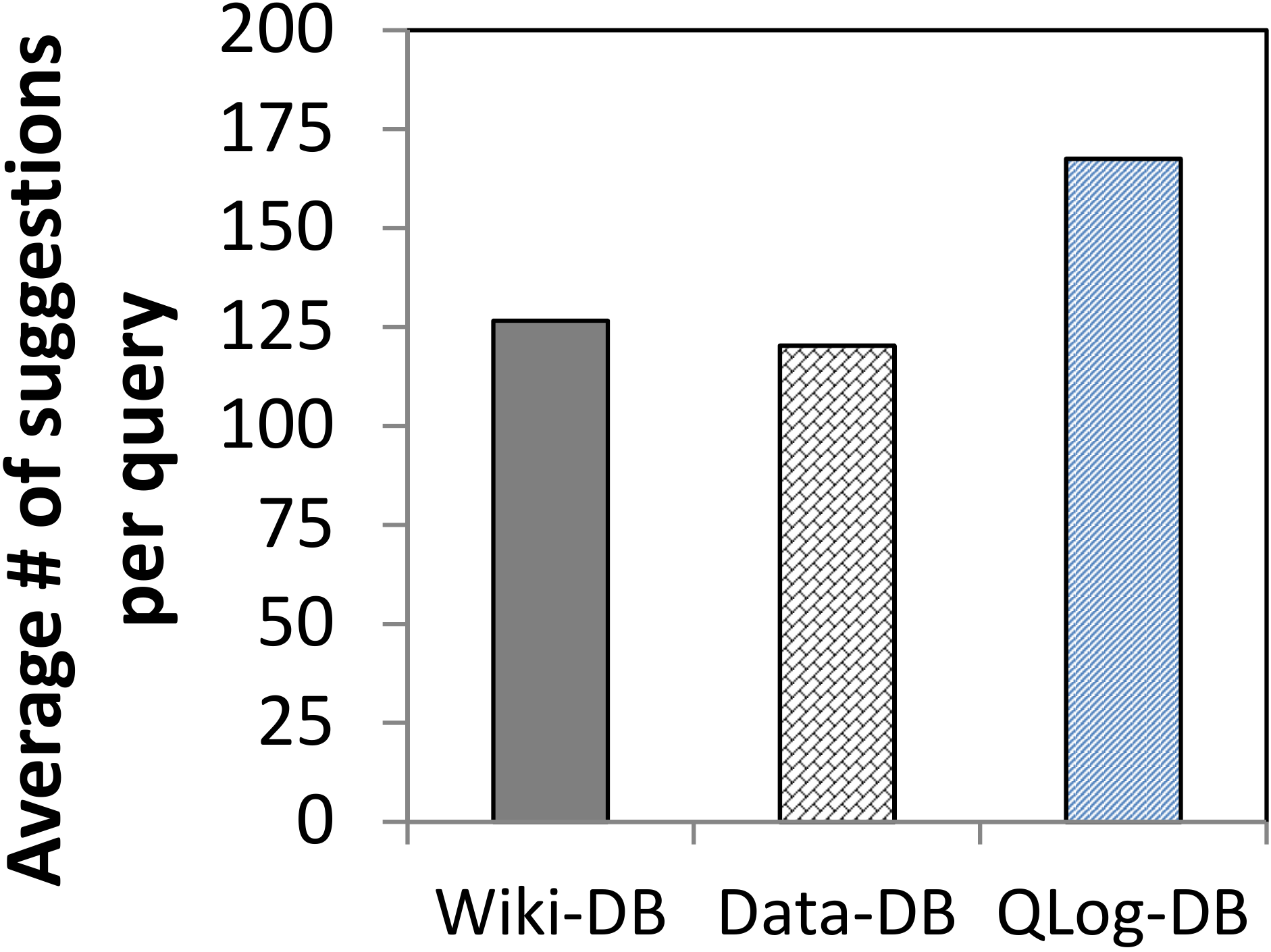}\vspace{1mm}
\label{fig:algo-log-db}
}
\vspace{-3mm}
\caption{Effectiveness of Query Logs}
\label{fig:algo-log}
\end{minipage}\vspace{-3mm}
\begin{minipage}[b]{0.55\linewidth}
\centering
\hspace{2mm}
\subfigure[Freebase]{
\includegraphics[width = 0.465\linewidth, angle=360]{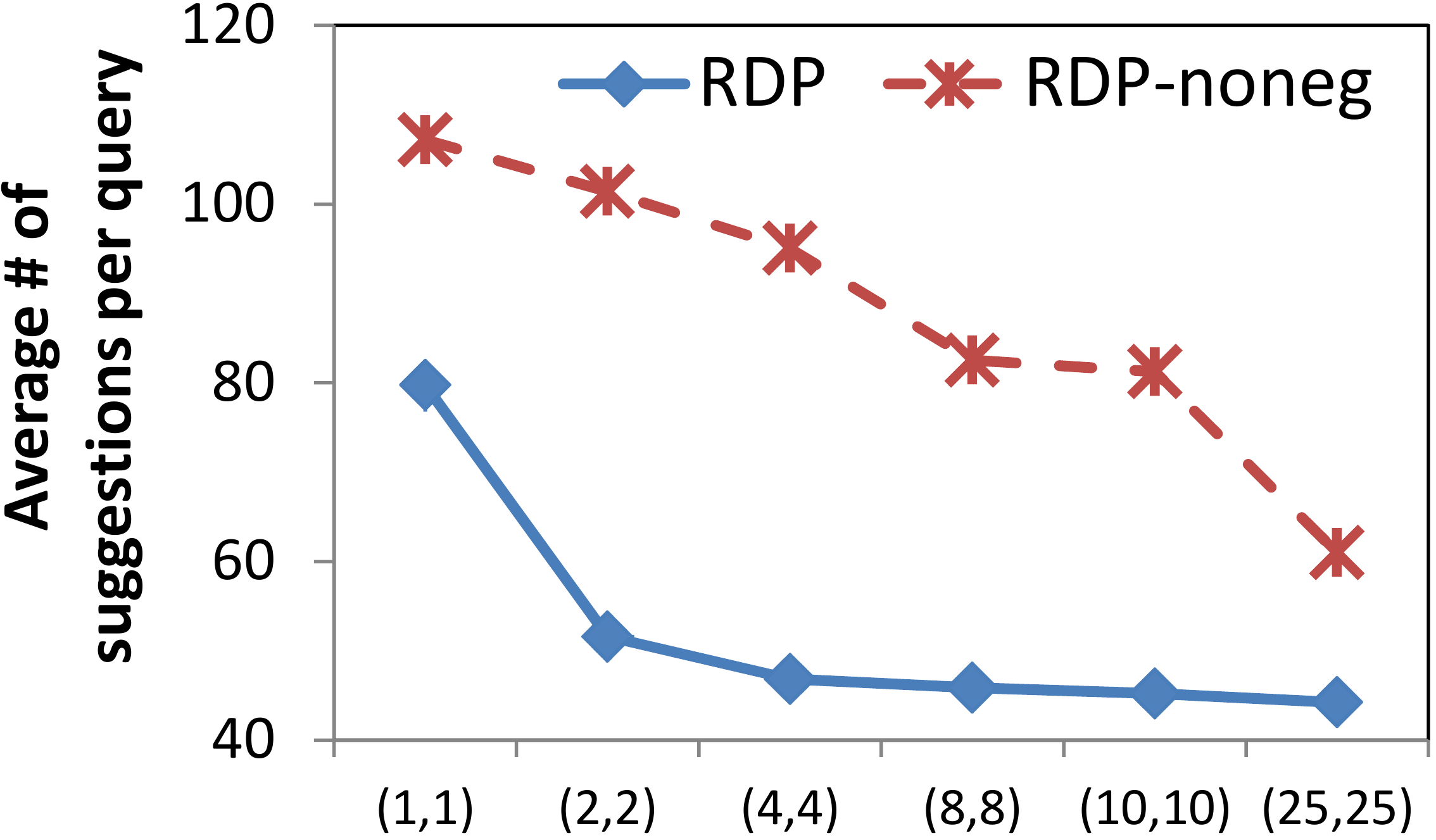}\vspace{1mm}
\label{fig:algo-param-fb}
}
\subfigure[DBpedia]{
\includegraphics[width = 0.465\linewidth, angle=360]{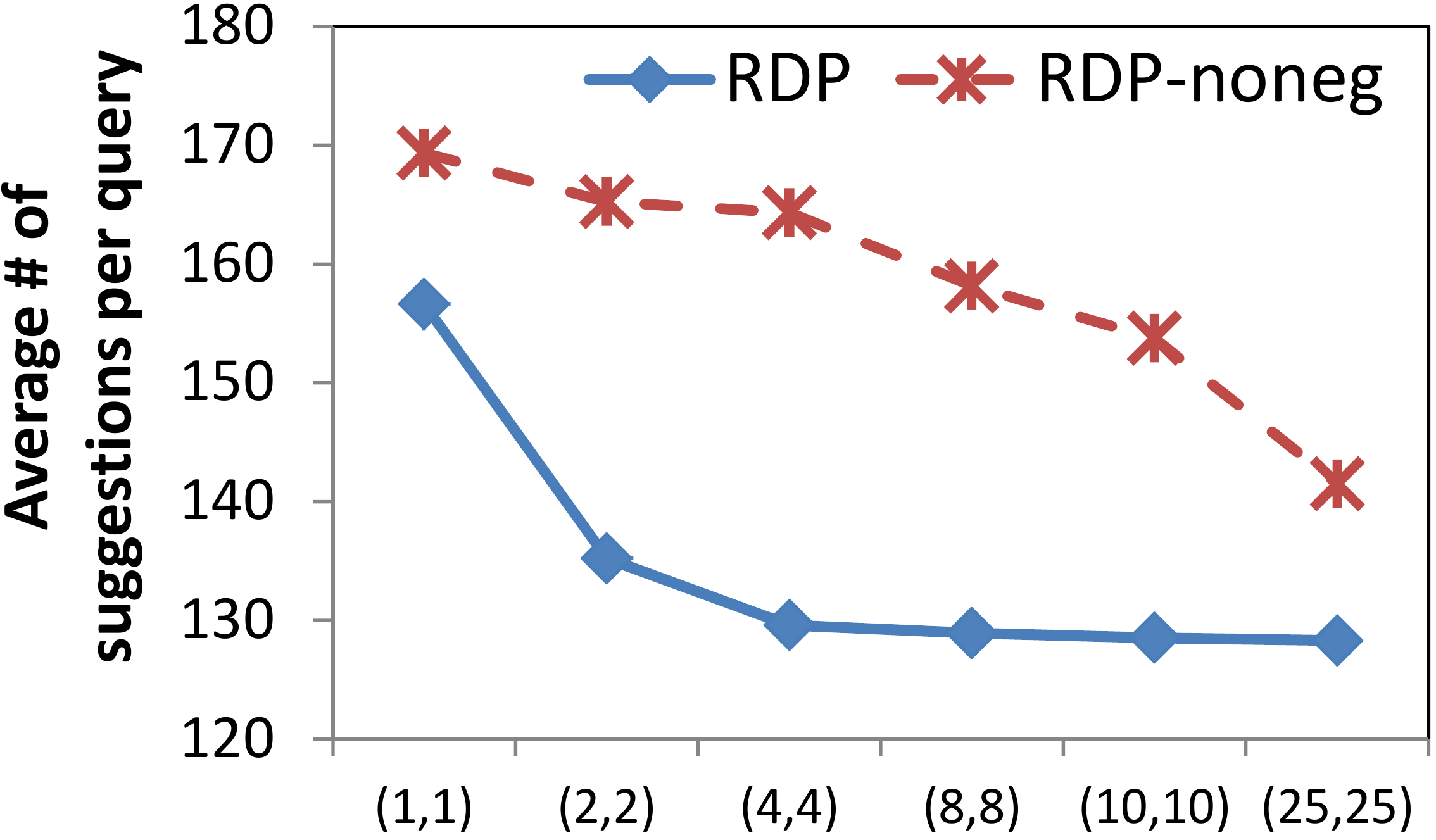}\vspace{1mm}
\label{fig:algo-param-db}
}
\vspace{-3mm}
\caption{Effect of Parameters on RDP ($N$, $\tau$)}
\label{fig:algo-param}
\end{minipage}\vspace{-3mm}
\end{figure*}

\subsection{Comparing Candidate Edge Ranking Methods}\label{sec:algoscompare}
We next compare the performance of RDP, \system{Orion}'s edge ranking algorithm, with other
machine learning algorithms: RF, NB, SVD and CAR. We compared
the performance of these algorithms over two widely used real-world data graphs: Freebase and DBpedia.
We used the Wiki-FB and Wiki-DB query logs for Freebase and DBpedia respectively.
We had to perform these experiments on the TACC machine, because RF has high memory requirements.
For instance, generating a random forest model with 80 trees, using a query log containing around 100,000 query sessions, requires 55 GB of RAM.

We created multiple target query graphs for each dataset, conforming with the schema of the
underlying data graph. For a given target query graph,
the input to each of the algorithms was an initial partial query graph containing exactly one edge
in it. The task of each algorithm was to iteratively suggest exactly one edge at a time, given the
partial query graph. If the edge suggested was present in the target query graph, it was added
into the partial query graph, and recorded as a positive edge. If not, the edge was ignored,
and recorded as a negative edge. The process was stopped either when the partial query graph was grown
completely into the target query graph, or if 200 suggestions were up. For each target query graph $G_t$
containing $E(G_t)$ number of edges, we internally converted it into $E(G_t)$ different instances
of target query graphs, each starting with a different-edged initial partial query graph as input to
the algorithms.

We created 43 target query graphs for Freebase, consisting of 6 two-edged query graphs,
10 three-edged query graphs, 9 four-edged query graphs, 17 five-edged query graphs and 1 six-edged
query graph. These 43 target query graphs were thus converted to 167 different input instances, creating
a query set called \emph{Freebase-Queries}. We created 33 target query graphs for DBpedia,
consisting of 2 three-edged query graphs, 29 four-edged query graphs, and 2 five-edged query graphs.
These 33 target query graphs were converted to 130 different input instances, creating a query set
called \emph{DBpedia-Queries}.

\subsubsection{Efficiency Based on Number of Suggestions}
For a query graph completion system, we believe an important
measure of its efficiency is the number of suggestions required to successfully grow a partial query
graph to its corresponding target query graph. This is because, if a system can help users construct
the target query graph with fewer number of suggestions, it indicates that the suggestions made indeed
captured the user's query intent. Figure~\ref{fig:algo-iters-fb} shows the average number
of suggestions required to complete each of the 167 input instances for Freebase. We observe that
RDP significantly outperforms the other methods. RDP requires only 43.5 suggestions per query graph
on average, nearly half the number of suggestions required to complete a query graph using RF and NB.
It also requires only a quarter of the number of suggestions required to complete a query graph
using SVD, while CAR requires 67.8 suggestions. Figure~\ref{fig:algo-iters-db} shows the average
number of suggestions
required to complete each of the 167 input instances for DBpedia. We observe that RDP requires
126.6 suggestions on average to complete a query graph, performing slightly better than NB which
requires 134.3 suggestions. RDP also comfortably outperforms RF, SVD and CAR which
on average require 164, 150.7 and 157.9 suggestions per query graph respectively.

\subsubsection{Efficiency Based on Time}
We next compare the efficiency of the various methods over the time required to grow the
initial partial query graph to its corresponding target query graph.  Figure~\ref{fig:algo-time-fb}
compares the average time required to complete a query task by each of the algorithms
over Freebase. RDP, NB and RF significantly outperform SVD and CAR.
RDP requires 7.7 seconds, slightly higher than NB's 3.9 seconds, and better than RF's 11.8 seconds
per query, which is commendable especially since both random forest and
Bayesian classifiers are extremely efficient once the models are learnt. Figure~\ref{fig:algo-time-db}
compares the average time required to complete a query task by each of the algorithms
over DBpedia. SVD and CAR are inefficient requiring 250.2 and 444.2 seconds per query respectively.
NB requires 5.9 seconds, which is faster than both RF and RDP that require 26.7 and 119.7 seconds per
query respectively.

\subsection{Effectiveness of Query Logs}
We compare the effectiveness of the various query logs listed in Table~\ref{tab:querylogs}. We use
RDP as the algorithm for edge suggestion, and the number of suggestions required to grow the initial
partial query graph to the target query as the measure of effectiveness of the query logs.
Freebase-Queries and DBpedia-Queries, described in Section~\ref{sec:algoscompare}, were the sets of
queries used to compare the various Freebase and DBpedia query logs respectively.

\textbf{Query Logs for Freebase:}\hspace{2mm}
Figure~\ref{fig:algo-log-fb} shows the
distribution of the number of suggestions required to complete a query task using Wiki-FB and Data-FB query
logs. We observe that 83 of the 167 input instances needed no more than
26 edge suggestions with the Wiki-FB query log, while it required at most 65 edge suggestions
to complete the same number of queries using the Data-FB query log. Around
42 more input instances required between 26 to 47 suggestions with Wiki-FB, while it required
between 65 to 200 suggestions with Data-FB. This indicates that the query log simulated using
Wikipedia and the Freebase data graph using \system{WikiPos} described in Section~\ref{sec:workload} is of
superior quality compared to the one simulated using only the Freebase data graph.
This suggests that positive edges established based on the context of human usage of the
relationships is better than the positive edges established using only the data graph.

\textbf{Query Logs for DBpedia:}\hspace{2mm}
Figure~\ref{fig:algo-log-db} shows the average number of edge
suggestions required to process the 130 different DBpedia input instances, using each of the
three aforementioned query logs for DBpedia. We first observe that QLog-DB performs poorly compared
to the other two query logs. This is because the DBpedia SPARQL query log is not comprehensive
enough and is limited in the variety of relationships captured, making it ineffective.
The second interesting
observation we make is the algorithm requires 120.3 suggestions on average using Data-DB,
while it requires 126.6 suggestions with Wiki-DB. Data-DB performs
slightly better than Wiki-DB due to the fact that DBpedia is a high quality data graph
generated using the info-boxes in Wikipedia pages. The sets of positive edges in
Wiki-DB are simulated using the text in Wikipedia and the DBpedia data graph. The
two query logs are thus highly similar to each other, unlike the case in Freebase where we could
see a significant difference between the performance of Wiki-FB and Data-FB.

\subsection{Parameter Tuning for RDP}
We finally study a variation of RDP, and the effect of $N$ and $\tau$,
the two parameters used in RDP.
As described in Section~\ref{sec:rcp}, given a query session $Q$, RDP builds $N$ different random
decision paths. Each random decision path is grown incrementally, until either the support for
the path is no more than a threshold $\tau$, or if all edges in $Q$ are exhausted. While building a
random decision path, RDP considers both the positive and negative edges. To study if considering
the negative edges indeed helps in better identifying the user's query intent, we create a variation
of RDP, called RDP-noneg, which does not include any negative edges in the random decision paths.
Figures~\ref{fig:algo-param-fb} and~\ref{fig:algo-param-db} compare the average number of suggestions
required to complete each query graph with different values of $N$ and $\tau$, for Freebase and DBpedia
queries respectively. In both the cases, we observe that the average number of suggestions required per
query decreases as we increase the number of random decision paths, and the threshold $\tau$.
It saturates after we reach around 10 for both $N$ and $\tau$ in RDP.
Figures~\ref{fig:algo-param-fb} and~\ref{fig:algo-param-db} also compare the average number of suggestions
required to complete the query graphs using RDP and RDP-noneg. With the best parameter values of $N=25$
and $\tau=25$, RDP requires
44.2 suggestions while RDP-noneg requires 60.9 suggestions in Freebase. RDP also requires fewer
suggestions in DBpedia with 128.5 suggestions compared to 141.5 suggestions required by RDP-noneg.
We observe that RDP significantly outperforms its variation RDP-noneg, indicating that considering
negative edges in query sessions is indeed helpful.

\section{Conclusions}
We introduce \system{Orion}, a visual query builder that helps schema-agnostic users construct complex
query graphs by automatically suggesting new edges to add to the query graph. \system{Orion}'s edge
ranking algorithm RDP, ranks candidate edges by how likely they will be of interest to the user,
using a query log. Since there are no real-world query logs, we propose several ways of simulating
a query log. User studies show that \system{Orion} has a 70\% success rate of building complex query
graphs, significantly better than a baseline system resembling existing visual query builders,
that has a 58\% success rate. We also compare RDP with several methods based on other machine learning
algorithms and observe that, on average, those other methods require 1.5-4 more suggestions to complete query graphs.

\begin{small}
\bibliographystyle{abbrv}

\end{small}
\end{document}